\documentclass{article}

\usepackage{arxiv}

\usepackage[utf8]{inputenc} 
\usepackage[T1]{fontenc}    
\usepackage{hyperref}       

\usepackage{url}            
\usepackage{booktabs}       
\usepackage{amsmath}
\usepackage{amsfonts}       
\usepackage{amssymb}
\usepackage{bm}
\usepackage{nicefrac}       
\usepackage{microtype}      
\usepackage{lipsum}
\usepackage{graphicx}
\usepackage{soul}
\sethlcolor{yellow}
\usepackage{xcolor}
\usepackage{amssymb,amsfonts,mathtools}
\usepackage{bm}
\usepackage{enumitem}
\usepackage{amsthm}
\usepackage[normalem]{ulem}           
\usepackage{cancel} 
\usepackage{cleveref}       
\usepackage{subcaption}

\usepackage{hyperref}

\hypersetup{
  pdftitle={
    If the Brain Were So Simple:
    Information-Based Drivers of Intelligence and Their Interaction
    Through a Proposed N-Body-Inspired Type Framework
  },
  pdfauthor={
    Brett J. Kagan, Johnson Zhou,
    Forough Habibollahi, and Valentina Baccetti
  },
  pdfsubject={
    A theoretical perspective on information enthalpy,
    neural information processing, and interacting neural drivers
  },
  pdfkeywords={
    information enthalpy, predictive information,
    neural computation, free-energy principle,
    active inference, neural criticality
  }
}

\hypersetup{colorlinks=true,linkcolor=black,citecolor=blue,urlcolor=black}
\usepackage[numbers,sort&compress]{natbib}
\newcommand{\E}{\mathbb{E}}

\newcommand{\RR}{\mathbb{R}}

\newcommand{\IEP}{\mathrm{IEP}}

\newcommand{\bbN}{\mathbb{N}}
\newcommand{\Mn}[1]{M_{#1}^{(n)}}
\newcommand{\Mntau}[1]{M_{#1}^{(n),\tau}}
\newcommand{\cMn}{\mathcal{M}^{(n)}}

\newcommand{\Cst}[1]{C_\mathrm{st,#1}}  
\newcommand{\Tcap}[1]{T_\mathrm{cap,#1}}  
\newcommand{\Cr}[1]{C_\mathrm{r,#1}}  
\newcommand{\Rt}[1]{\mathcal{R}^{(n)}_{#1}} 
\newcommand{\Rst}[1]{\mathcal{R}^{(n)}_{\mathrm{st},t}} 
\newcommand{\Ru}[1]{\mathcal{R}^{(n)}_{\mathrm{use},t}} 

\DeclareMathOperator{\MS}{MS}

\newcommand{\Hsym}{H_{\mathrm{sym}}}

\newcommand{\HI}{\mathcal{H}_I}
\newcommand{\PhiFun}{\Phi_\pi }

\graphicspath{ {./images/} }

\newcount\eqtagmode
\usepackage{tikz}
\usetikzlibrary{tikzmark, fit, backgrounds}
\usepackage{soul}

\makeatletter
\newcommand{\tagequations}{\@ifstar{\tagequations@text}{\tagequations@auto}}

\newcommand{\tagequations@auto}[3][]{%
    \ifcase\eqtagmode
        #3%
    \or
        \ifmmode
            \tikz[baseline, remember picture]{
                \node[anchor=base, inner xsep=4pt, inner ysep=2pt, rounded corners=4pt, fill=#2, #1]
                {$\displaystyle #3$};%
            }
        \else
            \smash{\tikz[baseline, remember picture]{%
                \node[anchor=base, inner xsep=4pt, inner ysep=1.2pt, rounded corners=4pt, fill=#2, #1]
                {$\displaystyle #3$};%
            }}
        \fi
    \or
        \ifmmode
            \tikz[baseline, remember picture, inner sep=0pt, outer sep=0pt]{%
                \node[anchor=base, inner sep=0pt, outer sep=0pt, trim left=0pt, trim right=0pt, #1] (hl) {$\displaystyle #3$};%
                \begin{pgfonlayer}{background}%
                    \draw[line width=3pt, color=#2]([yshift=1pt]hl.south west) -- ([yshift=1pt]hl.south east);%
                \end{pgfonlayer}%
            }
        \else
            {
                \setulcolor{#2}%
                \setul{0pt}{3pt}%
                \smash{\ul{\xspace#3}}%
            }
        \fi
    \fi
}

\newcommand{\tagequations@text}[3][]{%
    \ifcase\eqtagmode
        #3
    \or
        \smash{\tikz[baseline, remember picture]{%
            \node[anchor=base, inner xsep=4pt, inner ysep=1.2pt, rounded corners=4pt, fill=#2, #1]
            {$\displaystyle #3$};%
        }}
    \or
        {
            \setulcolor{#2}%
            \setul{0pt}{3pt}%
            \smash{\ul{\xspace#3}}%
        }
    \fi
}

\definecolor{color_encoding_kernel}{RGB}{255,181,112} 
\definecolor{color_stochastic_states}{RGB}{248,206,203} 
\definecolor{color_states_past}{RGB}{214,239,224} 
\definecolor{color_memory_current}{RGB}{235,210,250}
\definecolor{color_memory_past}{RGB}{180,200,255}
\definecolor{color_predictve_resource}{RGB}{216,226,220}
\definecolor{color_storage_process}{RGB}{114,10,183}
\definecolor{color_observations}{RGB}{82,182,154}
\definecolor{color_internal_states}{RGB}{153,217,139}
\definecolor{color_actions}{RGB}{217,237,146}
\definecolor{color_states}{RGB}{253,201,32}
\definecolor{color_raster}{RGB}{217,231,238}
\definecolor{color_patch}{RGB}{214,236,229}
\definecolor{color_symbol}{RGB}{232,222,235}
\definecolor{color_symbols}{RGB}{242,220,215}

\DeclareRobustCommand{\tagEncodingKernel}{\@ifstar{\tagequations@text{color_encoding_kernel!70}}{\tagequations@auto{color_encoding_kernel!70}}}
\DeclareRobustCommand{\tagStochasticStates}{\@ifstar{\tagequations@text{color_stochastic_states!70}}{\tagequations@auto{color_stochastic_states!70}}}
\DeclareRobustCommand{\tagStatesPast}{\@ifstar{\tagequations@text{color_states_past!70}}{\tagequations@auto{color_states_past!70}}}
\DeclareRobustCommand{\tagMemoryCurrent}{\@ifstar{\tagequations@text{color_memory_current!80}}{\tagequations@auto{color_memory_current!80}}}
\DeclareRobustCommand{\tagMemoryPast}{\@ifstar{\tagequations@text{color_memory_past!80}}{\tagequations@auto{color_memory_past!80}}}
\DeclareRobustCommand{\tagPredictiveResource}{\@ifstar{\tagequations@text{color_predictve_resource!80}}{\tagequations@auto{color_predictve_resource!80}}}
\DeclareRobustCommand{\tagInfoStorageProcess}{\@ifstar{\tagequations@text{color_storage_process!20}}{\tagequations@auto{color_storage_process!20}}}
\DeclareRobustCommand{\tagObservations}{\@ifstar{\tagequations@text{color_observations!40}}{\tagequations@auto{color_observations!40}}}
\DeclareRobustCommand{\tagInternalStates}{\@ifstar{\tagequations@text{color_internal_states!40}}{\tagequations@auto{color_internal_states!40}}}
\DeclareRobustCommand{\tagActions}{\@ifstar{\tagequations@text{color_actions!40}}{\tagequations@auto{color_actions!40}}}
\DeclareRobustCommand{\tagStates}{\@ifstar{\tagequations@text{color_states!40}}{\tagequations@auto{color_states!40}}}
\DeclareRobustCommand{\tagRaster}{\@ifstar{\tagequations@text{color_raster}}{\tagequations@auto{color_raster}}}
\DeclareRobustCommand{\tagPatch}{\@ifstar{\tagequations@text{color_patch}}{\tagequations@auto{color_patch}}}
\DeclareRobustCommand{\tagSymbol}{\@ifstar{\tagequations@text{color_symbol}}{\tagequations@auto{color_symbol}}}
\DeclareRobustCommand{\tagSymbols}{\@ifstar{\tagequations@text{color_symbols}}{\tagequations@auto{color_symbols}}}
\makeatother

\definecolor{applegreen}{rgb}{0.55, 0.71, 0.0}

\title{If the brain were so simple: Information-based drivers of intelligence and their interaction through an n-body-inspired framework.}

\author{
 {\large Brett J. Kagan$^{1}$\thanks{Corresponding author: brett@corticallabs.com}\;, \, 
 Johnson Zhou$^{1}$, \,
 Forough Habibollahi$^{1}$, \,
  Valentina Baccetti$^{1}$}
   \\[2ex]
$^{1}$Cortical Labs, Melbourne, 3000, VIC, Australia.\\[0.5ex]
}

\begin{document}

\maketitle

\vspace{20pt}

\begin{abstract}
Theories of how the brain processes information and returns intelligent outputs are numerous and often difficult to conclusively test. Here, we consider an approach grounded in informatics and algorithmic thermodynamics for how neural systems respond to different patterns of information in different ways, with a focus on the relationship between the information entropy of a signal and a new proposed quantity we term \textit{information enthalpy}, which represents the internal resource of structured information within a system available to be used for predictions and other forms of work. To evaluate the capacity of the external signals to increase information enthalpy in the system, we introduce the \textit{information enthalpy potential} (IEP) as a defined metric. We propose and implement a method for quantifying the amount of potential information enthalpy - the IEP - in a given signal across a range of example signals. We offer a conjecture of how information enthalpy may be treated by neural systems within a biological framework before showing how it may integrate and offer falsifiability to existing approaches such as the Free Energy Principle. This framework is also positioned in light of the role of neural criticality. Finally, we postulate a testable and falsifiable framework where the distinct mechanisms within each driver interact through a described n-body-inspired model to govern the ``motion'' of a neural system through a high dimensional state-space. Through these processes, it is proposed that these features form the most fundamental basis of the neural drivers that give rise to the complexity of adaptive behaviors that are commonly called: intelligence.   

\end{abstract}

\section{Introduction} 
\label{Introduction}  

How intelligence arises has been a question predating modern sciences, appearing in the earliest philosophical speculation. Countless hypotheses, theories, and principles have been proposed, from foundational Hebbian plasticity mechanisms \cite{zenke_hebbian_2017} to hierarchical multiscale models \cite{robinson_multiscale_2005} and system-wide principles for optimization \cite{friston_free-energy_2010, Trapp_2025_theory}. However, an integrated theory that ties the overlapping structural properties with the known functional properties in a physiologically plausible manner has never been formalized. As neurotechnology advances with the rise of systems such as in vivo brain-computer interfaces (BCI) and in vitro synthetic silico-biological intelligence (SBI) devices, it becomes vital to develop more advanced theories. Here, we outline a testable and falsifiable theory of how intelligence arises through an interaction of information processing drivers that seeks to integrate multidisciplinary perspectives and account for the variety of phenomena observed in neural structure and function. Importantly, we do not seek to reinvent and discard existing theories where evidence is supportive, instead the focus here is to integrate these theories in a coherent framework.  

\subsection{Information-based drivers of information processing and intelligence}

Investigations of intelligence and learning, whether biological or machine, have largely focused on reward mechanisms. In computer science this typically involves a reward function while in biology the focus is often on key neurotransmitters, such as dopamine~\cite{schultz_predictive_1998}. While dopamine and other neurotransmitters undoubtedly play a critical role in intelligence and learning in the biology we observe today\cite{previc_dopamine_1999, Goulty_2023_monoaminergic}, these helpful chemical messengers were not always evolved or indeed necessary for simpler action sets, where even bacteria or Hydra are able to display basic associative learning~\cite{bennett_five_2021, hanson_being_2023, van_duijn_phylogenetic_2017, Brette_2021_Integrative}. So while neurotransmitter-driven plasticity is certainly a critical "shortcut" for learning and intelligence functions, at the most fundamental level, it remains reasonable to ask whether the earliest origins of intelligence and the capacity to learn might have different drivers that are not wholly dependent on these chemical messengers. 

Here we explore the concept that information (perceived sensation with a given structure) is processed differently by biological neural systems based on the structure of this information. This idea is, of course, self-evident when discussing reactionary behavior (i.e. fixed action patterns) as even in insects complex actions (such as striding and mating behavior) can be elicited from specific structured stimulation \cite{ronacher_innate_2019, Buschges_2025_Motora}. However, where specific actions are not tightly controlled, do more general behaviors emerge in response to broader structural properties of sensory information?

Under this paradigm, the processing of a specific type of structured information does not, in itself, act as a reward or a punishment, but functions as different biological imperatives that facilitate the dynamic system to undertake more complex behaviors. Where these behaviors are necessary for survival, the response of the neural systems to the structure of information allowed evolutionary persistence  and the opportunity for further phylogenetic refinement (evolution of the system), which ultimately gave rise to more complex behavior, facilitated by yet more complex biological machinery~\cite{cisek_resynthesizing_2019}. 

Yet underpinning and preceding these traits, the most basic properties of information may reasonably be proposed to have been determined, or at least significantly influenced, by base physics and thermodynamic laws. That a core function of even the most rudimentary neural systems was to interact with the external environment is a sensible premise\cite{richardson_intelligence_nodate}. As shared physical laws exist as a property of the external environment that neural systems must exist within, this would accord with the proposed convergent evolution of neural systems that show similar responses to similar information patterns \cite{liebeskind_complex_2016, bucher_evolution_2015}. One field that has laid the groundwork for this approach is algorithmic thermodynamics, which extends work in classical thermodynamics to describe information as a physical quantity with given properties \cite{baez_algorithmic_2013}. A key focus of work in algorithmic thermodynamics is the quantification of information entropy or alternatively, of negentropy (as tendency towards order in a system) \cite{ebtekar_foundations_2025}. However, other work has explored what properties of information might be considered attractive, motivating or beautiful to neural systems and these features are not well captured by the simple relationship between order and disorder alone \cite{schmidhuber_formal_2010, schmidhuber_simple_2007, aaronson_quantifying_2014}. Instead, the search for novel and interesting - as judged by certain complexity measures - patterns of information relative to previously observed patterns of information is deemed a key driver \cite{schmidhuber_formal_2010, schmidhuber_simple_2007, aaronson_quantifying_2014}. These complexity measures attempt to capture a core feature of data that is \textit{interesting} to neural systems, which tends to be a balance between order and disorder, much in the same way as popular music throughout history is neither random noise nor exclusively a single oscillatory beat. These various perspectives accord well to adopting a bottom-up approach towards exploring how intelligence, learning and even cognition might arise, an approach which has had compelling advocation in the past \cite{de_waal_towards_2010}. The motivation behind this work is to explore different drivers that work to facilitate how a biological agent responds to different structures of information.

\subsection{Active Inference and Markov Blankets as Processes}

Active inference has been proposed as a unifying principle between action and perception, where behavior depends on the brain's internal models that predict future perceptions and govern actions accordingly \cite{pezzulo_active_2024, friston_active_2025}. As a concept, active inference has evolved from several directions, with the earliest roots going back to Helmholtz in 1866\cite{von_helmholtz_concerning_2001}, who proposed that perception is an unconscious inference based on an internal generative model with a series of recurrent top-down and bottom-up counter streams of processing of information \cite{pezzulo_active_2024}. Variations on this theme have arisen in psychology, cognitive science \cite{tolman_cognitive_1948}, cybernetics\cite{miller_plans_2017}, and computational approaches (such as the "Bayesian Brain hypothesis")\cite{doya_bayesian_2007}, as previously covered in depth ~\cite{pezzulo_active_2024, parr_active_2022}. Broadly active inference is cast as a normative framework shaping the behavior and cognition of living organisms \cite{parr_active_2022}. The \textit{active} component proposes that these organisms must adaptively control their action-perception loops to solicit a desired sensory observation (perception)\cite{parr_active_2022, parr_active_2017}. The \textit{inference} component involves a Bayesian generative model constructing an inferred state of the external world, particularly for which latent or hidden states give rise to the sensory observation, modeled through internal processes \cite{friston_active_2012, friston_active_2025}. Between these external and internal processes sits a statistical boundary which has been termed a \textit{Markov blanket} \cite{friston_life_2013}. A Markov blanket is that which separates any internal state from an external state and provides embodiment - in that a body separates the internal from the external - where the boundary comprises sensory and active states \cite{friston_active_2025, palacios_markov_2020}. In this way, a biological self-organizing system may generate and maintain Markov blankets at multiple scales and even in a nested manner. For example, a neuron has a Markov blanket defined by the cell membrane, as part of a network of cells with their own Markov blanket as defined by their topology, which rests inside a brain with its own Markov blanket as defined by the brain's geometry, which then sits inside a body. Such Markov blankets are intuitive, however, it has also been argued that biophysical boundaries are not a hard requirement for a Markov blanket to be formed \cite{kirchhoff_markov_2018}. Regardless of the exact boundaries in which Markov blankets do exist (a topic far beyond the scope of this paper), these Markov blankets are maintained by the system as a form of autopoietic organization \cite{palacios_markov_2020}, where the interaction between system and environment is mediated by the boundary that acts to effect an informational control of the environment under a certain policy \cite{auletta_paradigm_2010}. The principal normative framework proposed to govern these active-inference and Markov-blanket processes is the Free Energy Principle (FEP) \cite{kirchhoff_markov_2018, friston_free_2006}.

\subsection{The Free Energy Principle as a normative principle}

One feature of biological systems is that they maintain their states and form, despite a dynamic environment \cite{friston_free-energy_2010}.  One account on an inherent driver to modulate how a biological agent may respond to different structures of information would be to minimize the long-term average surprise of incoming sensation. This is done to ensure that sensory entropy remains low, thereby meeting the imperative of maintaining state and form and, consequently, homeostasis \cite{friston_free-energy_2010}. 
To this effect, the FEP posits that any self-organizing system that maintains a steady state (survives) must internally minimize a quantity called the variational free energy, which provides an upper bound on the system's self-information (also called surprise) associated with its sensory inputs \cite{friston_free-energy_2009}. Under this model, the organism acts upon sensory input based specifically on the internal model \cite{buckley_free_2017}. In theory there are two (non-exclusive) ways variational free energy can be minimized: 
\begin{enumerate}
    \item an organism can change its actions to control the external environment as to make future perception predictable; \item or it may get better at predicting the external environment regardless of action. 
\end{enumerate}  
Here is where active inference framework acts as a process driven by the FEP as a normative principle \cite{friston_active_2017}. Specifically, the active inference process operates by  sampling of the available environment through sensation, then acting to suitably minimize the divergence between the internal predictions and the external observations \cite{Friston_2025_Active}. Therefore, from a mathematical standpoint, the FEP specifically concerns the optimization of approximate posterior beliefs about hidden states of the environment to minimize their divergence from the distribution of sensory observations.
It is useful to consider how information entropy is defined in the FEP literature. For given sensory data $y$, and a model $E$ for how it was created, the information entropy is:
\begin{equation}
    H(y) = -\int P(y \mid E) \log P(y \mid E) \, dy
    \label{eq:internalInfoEntropy}
\end{equation}
where effectively the information entropy of the generalized sensory states is the average surprise (negative log probability of an outcome). In practice, organisms have an upper bound on surprise as sensory states must remain within physiological bounds. Likewise, because surprisal, \(-\log p(o)\), is generally not directly accessible to the system, the FEP introduces variational free energy as a tractable upper bound on surprisal. This upper bound has been proposed to be a key feature of self-organizing systems \cite{palacios_markov_2020}. The variational free energy term ($F$) is defined as:
\begin{equation}
    F[q] = \mathbb{E}_{q(s)}\big[\ln q(s) - \ln p(o,s)\big]
    \label{eq:VFE}
\end{equation}
 where $o$ are the observations (perceptions), $s$ are the hidden or latent causes of the observations, $q$ are the approximate posterior belief about hidden states and $p(o,s)$ is the generative model. Here $F[q]$ is an estimate based on what is currently observed (perceived through sensation) of how accurate a probabilistic explanation of hidden causes for that observation balanced against the complexity of that explanation. Yet when it comes to future-orientated action selection, the expected free energy ($G$) is used instead and defined as:
 \begin{equation}
 G(\pi)=\mathbb{E}_{q(o,s|\pi)}\big[\ln q(s|\pi) - \ln p(o,s|C)\big]
    \label{eq:EFE}
\end{equation}
where $\pi$ is the policy or selected actions, $q(o,s|\pi)$ is the predicted future outcomes and states under policy $\pi$, and $C$ is the prior preferences over outcomes. The $G(\pi)$ acts to set the expectation of the future uncertainty if a specific policy or set of actions is adopted. 

An important caveat is that the FEP often relies on an assumption called ergodicity. Ergodicity implies that the time a system spends in a particular state-space is equal to the probability that the system will be in that state-space when randomly sampled. In simple terms, this means that the states a system visits over time are assumed to reflect the states it is generally likely to occupy. This assumption helps connect the system's ongoing sensory experience with claims about its long-term average surprise or entropy \cite{colombo_non-equilibrium_2021}. Substantial literature has already explored the mathematical iterations and derivations of the FEP in detail, e.g. \cite{fields_free_2022, friston_active_2012, parr_active_2022, friston_free_2006, friston_free-energy_2010, buckley_free_2017, friston_free-energy_2009}, so this will not be repeated here. Likewise, the neuroscientific and psychological perspectives associated with the FEP specifically are equally broad and have been applied in multiple ways, e.g. \cite{Proietti_2025_Active, Trapp_2025_theory, parr_active_2017, parr_discrete_2018}. For example, at the computational level, the free energy minimization process has been approximated as a Bayesian inference process, while at a cellular level it has been cast as the neural population dynamics that encode predictions and prediction errors \cite{Proietti_2025_Active, kiebel_free_2011}. Fortunately, the above descriptions serve as a suitable foundation for positioning the concepts introduced in the rest of this paper. 

\subsection{Questions about FEP, information enthalpy, and neural criticality} 
Our previous work supported that altering the information entropy of a feedback signal following successful or unsuccessful gameplay can produce real-time learning in a biological neural network (BNN) \cite{kagan_vitro_2022}. Subsequent work further indicated that this learning was not reducible to simple pairwise synaptic plasticity alone, but involved population-level changes in network dynamics \cite{habibollahi_critical_2023, khajehnejad_dynamic_2025}. These findings are broadly compatible with the interpretations of the FEP that predicts biological systems should reduce prediction error or variational free energy through changes in perception, action, and internal states \cite{Friston_2025_Active}. However, they also expose a limitation in FEP when it is used as a standalone account of intelligent behavior. In particular, FEP explains why a system may move toward locally predictable states, but it does not by itself fully explain why a system should continue to seek complex, structured environmental input once local prediction error has been reduced. This limitation is often termed `the dark-room problem', which argues that if an organism or adaptive system is driven only to minimize surprise or variational free energy, then a simple strategy would appear to be withdrawal into an environment that is maximally predictable: a dark, quiet, unchanging room \cite{friston_free-energy_2012}. Of course such behaviors are seldom viewed as congruent with long-term survival. Early formulations addressing this challenge place the explanation on priors (usually genetic ones) whereby surprise is not defined relative to raw sensory entropy alone, but relative to the organism's generative model. Therefore a dark room is not genuinely unsurprising for an organism whose model expects movement, nourishment, social interaction, thermoregulation, and sensory engagement \cite{friston_free-energy_2012}. Active inference accounts follow a similar logic but extend it to expected free energy, epistemic value, and ambiguity reduction: agents do not merely minimize present prediction error, but select policies that reduce uncertainty about future states and expected outcomes \cite{schwartenbeck_exploration_2013, friston_active_2015, parr_generalised_2019}. However, these approaches depend on a generative model that is relatively complex. Whether rodents would display a capability for this level of higher order cognition is debatable, but it is implausible that simpler neural systems such as insects or in vitro neural cultures could display these traits. Yet these simple systems are still capable of diverse adaptive behaviors consistent with rudimentary intelligence\cite{cai_brain_2023, Robbins_2026_Goaldirected, kagan_vitro_2022, sumi_biological_2023, isomura_cultured_2015}.

A second potential FEP consideration is where free energy minimization is cast as precision in such a way as to explain complex intelligent behavior as a top down psychological control of an organism to maximize accuracy of predictions \cite{Trapp_2025_theory}. While compelling, this perspective does not explain bottom-up assembly of adaptive or rudimentary intelligent responses in the absence of intentional behavior. Similarly, this FEP perspective (if considered in isolation) presupposes a direct causal link between perception optimization and decision optimization without demonstrating that this is the case \cite{Trapp_2025_theory}. So while the FEP may lead a system to act to maximize the accuracy of its perception of a signal, both now and in the future, it does not explain why a system should prefer one type of signal over another, particularly with respect to signal complexity. Neither does this perspective answer why an organism would choose more complex signals over simpler signals. In practice, it may suggest the opposite as simpler signals may be easier to predict. Yet, humans and other organisms do have a preference for certain structures of information that generally have higher complexity, rather than simpler information structures \cite{schmidhuber_formal_2010, schaul_curiosity-driven_2011, schmidhuber_simple_2007}. This limitation interacts with the required assumption of ergodicity described above. Previously, the ergodicity assumption has been challenged because  dynamics of ensemble averages  are not generally equivalent to the dynamics of the individual systems, when those systems are not effectively decoupled from their history of interactions \cite{Aguilera_2022_How}. This challenge is particularly relevant when considering systems that learn and adapt, as ergodicity breaking behavior is effectively required for a system to transition and be maintained in new dynamic states consistent with memory and learning\cite{baccetti_ergodicity_2024}. As the core of intelligence involves adaptation, ergodicity persisting in an unstructured way through a system would be maladaptive.

Perhaps for this reason, while these FEP perspectives are interesting and useful, they do not present a complete framework to explain intelligent behaviors, especially at the most rudimentary level in simpler organisms. Critically, these features do not mean that FEP as a principle faces a direct challenge in what it sets out to do, rather we argue here that the FEP and active inference literature provides significant explanatory power to a wide range of phenomena. However, we would suggest that the drivers described in FEP alone cannot explain the breadth of intelligent behavior. Instead, we offer a hypothesis for a fundamental driver that interacts with the FEP, extending the theoretical tooling from a distinct perspective, and which would better explain these gaps in a testable and falsifiable way. We propose that biological neural systems are not only driven to reduce free energy, but also to maximize access to complex, structured information from their environment. We term this proposed driver \textit{information enthalpy maximization}. By \textit{information enthalpy}, we propose to describe an internal pool of structured information that has been taken up by a biological information-processing system and transformed internally into a usable resource. This should not be understood as a drive toward maximal entropy. Random noise is high entropy but poor in exploitable structure. Nor is it a drive toward minimal entropy, since a static signal provides little new information and rapidly collapses the opportunity for adaptive reorganization. Rather, information enthalpy refers to structured, temporally extended, non-random information that can be stored, integrated, and used by the information processing system to perform future information-requiring work. We propose that the term \textit{information enthalpy} is apt because (as we demonstrate below) there are conceptual parallels to thermodynamic enthalpy in a similar way that information entropy has conceptual parallels to thermodynamic entropy. Information enthalpy at a conceptual level is the sum of a system's internal structured information that is available to be used with a specific relationship (or exchange) to the surrounding environment and internal states and constraints on the system. Conveniently, the term also captures the intent for the concept to act as a counterweight to the free energy minimization process that involves information entropy. Conceptually, information enthalpy concerns the structured predictive information available to a system. In this work we distinguish between the realized predictive-information resource \(h_I\), representing information actually acquired and available for use, and the information-enthalpy capacity \(H_I\), representing the maximum realizable resource under the system’s constraints.     
 
An information enthalpy maximization driver offers a complementary, signal- and resource-oriented description that may generate operational predictions that better account for some of the above noted limitations not naturally foregrounded in standard FEP treatments. For example, a dark room may minimize sensory entropy, however, from the perspective of an information enthalpy-based driver, the dark room is impoverished. The dark room provides too little structured input to maintain the internal dynamical resources required for adaptive behavior. The system therefore has reason to leave the dark room not merely because it has inherited priors against darkness, nor only because it can model the epistemic value of future states, but because continued exposure to structured environmental information is necessary for maintaining its own capacity to process, store, and transform information.  In this sense, information enthalpy maximization drivers provide a more direct solution to the dark-room problem than FEP alone, along with addressing why systems prefer information structures that often align with features with particular complexity and structure, such as those called beauty. Information enthalpy as a driver would predict that intelligent systems should avoid both unstructured noise and perfectly predictable stasis, instead occupying environments that provide structured complexity. This proposal is supported by our observations that robust markers of neural criticality arise only when BNNs are embodied in simulated environments that provide structured information input \cite{habibollahi_critical_2023}. Criticality is commonly understood as a dynamical regime between order and disorder, rather than as a simple low-entropy state \cite{cocchi_criticality_2017}. This is significant because FEP, if interpreted primarily as a principle of uncertainty reduction, does not straightforwardly explain why a neural system should maintain itself near a critical regime. Critical systems are not maximally ordered, maximally predictable, or maximally quiescent. They exhibit rich fluctuations, large dynamic repertoires, and sensitivity to perturbation. These features are precisely what make criticality attractive as a necessary but not sufficient candidate for cognition (as previously suggested \cite{habibollahi_critical_2022, kagan_harnessing_2025})  but the same features are difficult to reconcile with a simplistic drive toward entropy minimization. Information enthalpy offers a natural explanation for this relationship. If structured environmental input supplies the informational resources required to maintain a critical state, then criticality is not merely an internal attractor of the network. It is an embodied relation between the network and the informational structure of its environment. Under this interpretation, the network maintains criticality by continuously sampling input that is complex enough to prevent collapse into order, but structured enough to avoid dissolution into noise. This also explains why less structured environments would be expected to produce less critical dynamics: they fail to provide the network with the informational gradients required to sustain an edge-of-chaos regime. Finally, if neural systems separate out their interactions with the world based on different drivers, it is reasonable that free-energy-minimization-based drivers work to establish ergodic state-sustaining relationships with the environment, while an informational enthalpy maximization drivers govern the non-ergodic interactions with the world. The competing interactions between these drivers and how they interact with a system aiming to maintain a critical regime would therefore give rise to the diverse and interesting phenomena that are broadly termed 'intelligence'. 


\section{Information Enthalpy}

The core concepts that this formalization aim to capture are: 1) how much `predictive' structure the system can hold about the type of information the system may receive in the future given a set of actions, and 2) the variables which modulate the rate at which this predictive structure (information) flows into and out of the system. 
In practice information enthalpy would be maximized in signals that have time-based information complexity. 
We can propose three ideas around this:
\begin{enumerate}
    \item Any useful information should come from the environment as sensory data perceived by the agent as established by the directionality of information.
    \item Any useful information should be stored by the agent in some structured way for later use (and may be captured in an active information storage framework as described below).
    \item The goal of the agent storing useful information should be to use it later for making predictions about the environment, and self-regulate.
\end{enumerate}

Information enthalpy is proposed to be a system property which is optimized over time through embodiment within an external environment in a closed loop.  To develop this framework, we first identify the key variables that must govern the relationship between a system and the environment. We start by considering time to be discrete, and represented by the time index $t\in\mathbb{Z}$. We consider the environment to have both hidden states $Z_t$, which are not directly detectable by the system, and observable states, represented as a $m$-dimensional set of real-valued vectors $O_t\in\RR^m$, which we assume constitute the system's sensory inputs. We assume that the sensory inputs interact with the system's internal neural states, represented by an $n$-dimensional set $X_t\in\RR^n$, and will facilitate actions $A_t\in\RR^p$, comprising $p$ components selected according to a given policy $\pi$. Finally, we assume that the system and the environment interact reciprocally at every time point $t$.

\begin{figure}[t]
    \centering
    \includegraphics[width=1.0\linewidth]{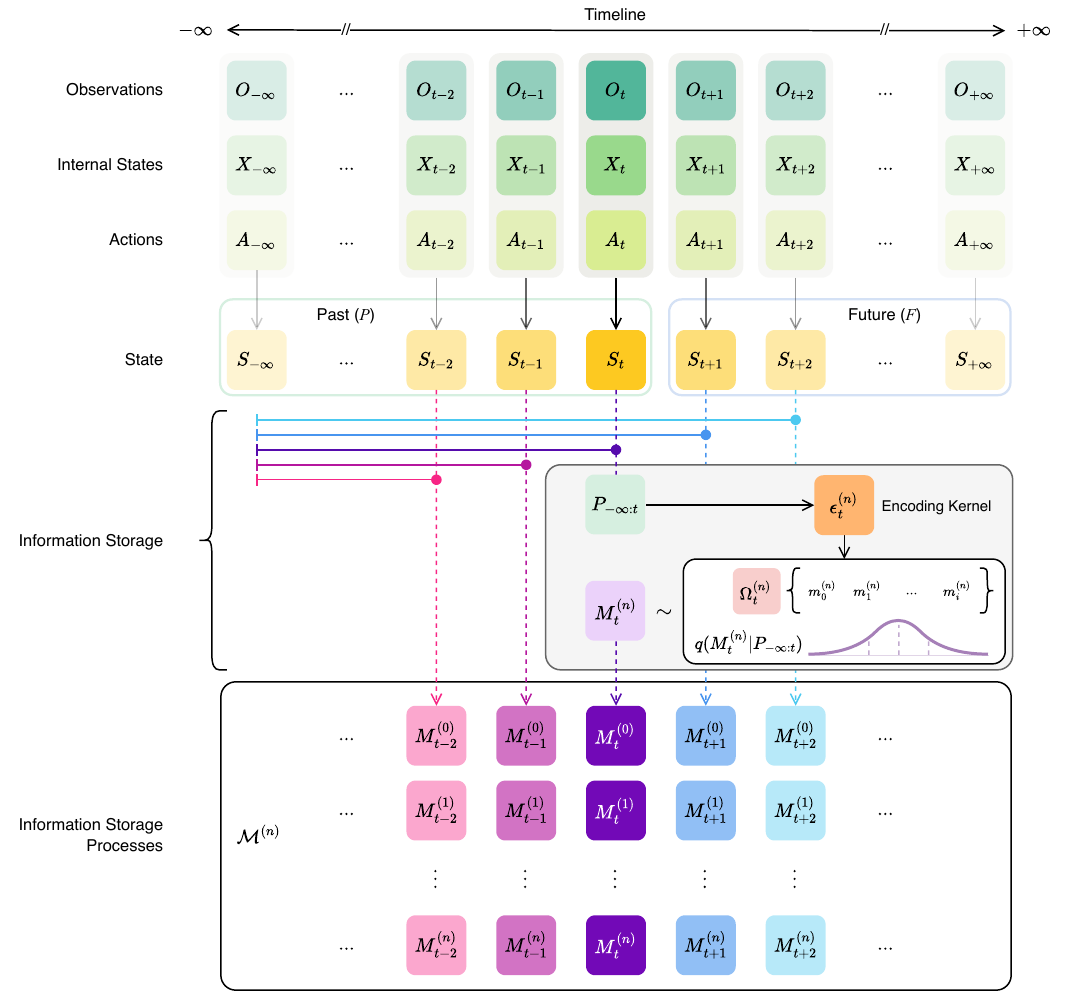}
    \caption{
    Schematic indicating the process of storing information from states $\tagStates{S_t}$, comprising of observations $\tagObservations{O_t}$, internal states $\tagInternalStates{X_t}$ and actions $\tagActions{A_t}$, to a set of $n$ information storage processes $\tagInfoStorageProcess{\mathcal{M}^{(n)}}$. For a single process, state $\tagMemoryCurrent{M^{(n)}_t}$ is sampled from a probabilistic set of possible states $\tagStochasticStates{\Omega^{(n)}_t}$ created by encoder $\tagEncodingKernel{\varepsilon^{(n)}}$ based on past states $\tagStatesPast{P_{-\infty:t}}$.
    }
    \label{fig:information_storage_schematic}
\end{figure}

\paragraph{Information-storage process.} Having established the variables describing the system and its interaction with the environment, we now introduce the temporal structure of these interactions. While the system is in a non-equilibrium state with respect to the environment, we assume that it shares a joint history with the environment through their closed-loop interactions. Therefore, the state at time point $t$ of both system and environment can be considered as influenced by a joint past $P_{-\infty:t}:=(O_{-\infty:t},X_{-\infty:t},A_{-\infty:t})$, and will go on to influence the joint future $F_{t+1:\infty}:=(O_{t+1:+\infty},X_{t+1:+\infty},A_{t+1:+\infty})$. The trajectory of the coupled environment-system ensemble is then represented by the time-series $S_{-\infty, +\infty} = (P, F)$, effectively mapping the shared past to the collective future (see~\Cref{fig:information_storage_schematic}). The trajectory $S_{-\infty, +\infty}$ may comprise both deterministic and stochastic components arising from the coupled system-environment dynamics, providing a common dynamical framework for the information-theoretic quantities introduced below.

To facilitate these interactions, the system is equipped with an internal information-storage mechanism that, at each time point $t$, encodes the joint interaction history $P_{-\infty, t}$ into an abstract information-storage representation. This is achieved through a causal stochastic encoding kernel\footnote{The present framework imposes no assumptions on the transition structure induced by the encoding kernel beyond causality. In particular, temporal regularity, including continuity or persistence of successive information-storage states, is regarded as a property of the specific encoder and of how the resulting representation is realized within the system’s internal states, rather than of the abstract information-storage framework.}
\begin{equation}
    \tagEncodingKernel{\varepsilon^{(n)}}: \tagStatesPast{P_{-\infty:t}} \to \Delta \left(\tagStochasticStates{\Omega^{(n)}_t}\right),
    \label{eq:encodingkernel}
\end{equation} 
where $\Omega^{(n)}_t$ denotes the state space of possible information-storage states, whose elements are $m^{(n)}_i \in \Omega^{(n)}_t$, \(\Delta (\Omega^{(n)}_t)\) the corresponding probability simplex, the superscript $(n)$ identifies the particular stochastic encoding kernel (encoder). For a given realized interaction history $P_{-\infty:t}$, the encoding kernel $\varepsilon^{(n)}$ assigns a probability distribution over the information-storage states $q(M^{(n)}_t\big| P_{-\infty:t})$ over the possible information-storage states in $\Omega^{(n)}_t$. Due to the stochastic nature of the encoder, the same interaction history may therefore generate different information-storage states.\footnote{Here stochasticity represents the inherent randomness and variability in how interaction histories are encoded and stored, consistent with the probabilistic nature of active inference.} The probability distribution  $q(M^{(n)}_t\big| P_{-\infty:t})$ produced by the encoding kernel defines the distribution of the information-storage state $\Mn{t}$. Accordingly, the information-storage state is obtained by sampling from the distribution
\begin{align}
    \tagMemoryCurrent{\Mn{t}} 
\sim 
     q(M^{(n)}_t\big| P_{-\infty:t}),
\end{align}

The sequence $\tagInfoStorageProcess{\cMn = \big(\Mn{t}\big)_{t\in\mathbb{Z}}}$ defines the corresponding  information-storage \textit{process}, comprising a temporal sequence of encoded representations of the system's past interactions. The collection $\mathfrak{M} = \{\mathcal{M}^{(1)}, \mathcal{M}^{(2)}, \ldots \}$ comprises the family of information-storage processes generated by all causal stochastic encoding kernels defined via Eq~\eqref{eq:encodingkernel}. Each process is associated with its own information-storage state space and representation of the interaction history. 
 
Each information-storage process $\cMn$ is uniquely determined by a single, time-invariant stochastic encoding kernel $\varepsilon^{(n)}$. Consequently, each information-storage state $\Mn{t}$ depends only on the interaction history available up to time $t$, ensuring causality. It is important to note that $\Mn{t}$ does not store the raw sequence of past sensory states $O_{-\infty:t}$, nor the raw joint interaction history $P_{-\infty:t}$. Rather, it is a processed or ``pooled'' representation of $P_{-\infty:t}$ that encompasses past observations, internal states, and actions. Each state $\Mn{t}$ should therefore be interpreted as an information pool containing an encoded representation of the system’s past interactions, from which information relevant to future operation can be extracted.

\paragraph{Storage of past states information within the system.} We assume that the information-storage states $\Mn{t}$ are realized within a subset $X'_t \subseteq X_t$ of the system's internal states dedicated to information storage, which we denote abstractly by $\Mn{t}=g(X'_t)$. This modeling assumption separates the subsystem responsible for storing $\Mn{t}$, from the remainder of the system states, which may support other computational or physiological functions.

The storage subsystem is characterized by a time-varying maximum information-storage capacity, denoted by  $\Cst{t}$,  which specifies the maximum amount of information that can be retained by the subsystem at time $t$. The inflow of information is measured via the throughput rate $\Cr{t}$ (or maximum rate at which information can be stored). We model its dynamics using the following difference equation 
\begin{align}
    \Cst{t}
=
    \big(
        1 
    -
        \alpha_t
    \big)
    \Cst{t-1}
+
    \Cr{t-1}
\label{eq:maxstorecapacity}
\end{align}
where $\alpha_t$ is the time-varying leaky parameter. while a more general derivation is provided in Appendix~\ref{appsec:T_cap}. 

The maximum storage capacity may increase through neural plasticity mechanisms, but would eventually approach an equilibrium with intrinsic decay, as detailed in sec.~\ref{sec:biocosts}, where the information accumulation  saturates concavely. It is important to notice that the leaky parameter term in Eq.~\eqref{eq:maxstorecapacity} describes the deterioration of the maximum storage capacity, not of the information stored. The rate of information storage also changes over time, due to plasticity. We model its dynamics as 
\begin{align}
    \Cr{t}
=
    u_t - v_t + \big(1 - \delta_{\mathrm{decay}} \big)\,\Cr{t-1}
\label{eq:storagerate}
\end{align}
where $u_t$, $v_t$, and $\delta_{\mathrm{decay}}$ are respectively the build rate, the prune rate, and the rate of capacity decay if not maintained.

The leaky parameter $\alpha_t$ establishes a natural time-scale $T_{\mathrm{cap},t} = 1/ \alpha_t$, that describes the \textit{historical capacity depth} the system can actively maintain (a time-varying parameter that controls the non-Markovian history depth)\footnote{This expression is dimensionally sound if we consider the definition of timepoints $t \in \mathbb{Z}$, for which $\Delta t = 1$. We will assume this to be the case throughout this article. The more general expression can be found in the appendix~\ref{appsec:T_cap}.}. Therefore, $T_{\mathrm{cap},t}$ is an effective, time-varying upper time bound:
\begin{equation}
    T_{\mathrm{cap},t}
=
    \max\!\Big\{t_\mathrm{cap},\ 
    \frac{
        \Cst{t-1}
        }
        {
        \Cr{t-1}
    -
        \left(
            \Cst{t}
        -
            \Cst{t-1}
        \right)
        }
    \Big\}
    ,
\label{eq:Hcap}
\end{equation}
Here, $t_\mathrm{cap}$ can be thought of as a minimum retention parameter that prevents the occurrence of a zero-horizon state $T_\mathrm{cap,t} = 0$ for the case $\Cst{t} \rightarrow 0$. This is particularly important in the context of active inference, where this retrospective window serves as the mandatory informational baseline for prospective action selection. Furthermore, as future policy optimization $\pi$ and expected information enthalpy maximization require conditioning on historical trajectories to preserve predictive consistency, the system’s predictive (future) horizon is also strictly bounded by its retrospective retention window. If this window were to be zero, the system's predictive capability would be drastically reduced (and would be hard to rebuild). This formulation creates a biologically plausible, state-dependent upper time bound based on the information already stored by the system for which a neural system can efficiently optimize the information-theoretic processes detailed in the following sections. 

From Eq.~\eqref{eq:Hcap} we can see that, to ensure that $\Tcap{t}$ is strictly positive, the storage capacity $\Delta \Cst{t}$ rate of change cannot exceed the throughput capacity rate $\Cr{t}$:
\begin{equation}
    \Delta \Cst{t}
=
    \Cst{t} - \Cst{t-1}
\le
    \Cr{t-1} ,
\end{equation}
in practice limiting that rate of change of $\Cst{t}$: the storage capacity cannot increase as fast as the maximum nominal rate at which information can be stored. It is important to notice that, this is due to the chosen dynamic behavior of $\Cst{t}$. 

To capture the chronological evolution of $\Cst{t}$ and $\Cr{t}$ across the system's entire operational lifespan, we formalize them as the complete trajectories $\mathbf{C}_{\mathrm{st}} \coloneqq (\Cst{t})_{t \in \mathbb{Z}}$ and $\mathbf{C}_{\mathrm{r}} \coloneqq (C_{\mathrm{r}, t})_{t \in \mathbb{Z}}$. However, note that for $T_{\mathrm{cap},t}$ to remain positive and finite, \(C_{r,t-1}-\Delta C_{\mathrm{st},t}\) must be strictly positive.

\paragraph{Stochastic variability of information storage state.} As described above, the causal stochastic kernel induces the conditional probability distribution $q(\Mn{t}|P_{-\infty:t}) $ over the set of admissible information-storage states $\Omega_t^{(n)}$.  The Shannon entropy of the resulting information-storage state is therefore given by
\begin{align}
    H(\Mn{t})
=
    -\sum_{m^{(n)}_i \in\Omega^{(n)}_t}
    p(m^{(n)}_i)
    \log
    p(m^{(n)}_i),
\end{align}
where $ p(m^{(n)}_i) = \Pr (\Mn{t}= m^{(n)}_i )$ is the probability that the information-storage state $\Mn{t}$ takes the value $m^{(n)}_i\in\Omega^{(n)}_t$.

The Shannon entropy $H(\Mn{t})$ quantifies the uncertainty of the stochastic information-storage state or, equivalently, the average information (in bits, nats, etc.) required to represent a realized information-storage state under an optimal encoding \cite{shannon_mathematical_1948,cover_elements_2006}. Since $\Mn{t}$ is realized within a storage subsystem of finite capacity, we assume that the average information required to represent an admissible information-storage state cannot exceed the available storage capacity,
\begin{align}
    H(\Mn{t})
\le
    \Cst{t},
\label{eq:entropyMcap}
\end{align}
so that the storage capacity $\Cst{t}$ provides an upper bound on the entropy of $\Mn{t}$.    

\paragraph{Redundant information.} While the entropy $H(\Mn{t})$ quantifies the average information required to specify the realized value of the information-storage state, it does not distinguish between information that is relevant for prediction, information that is statistically duplicated or otherwise unnecessary for predictive function, and variability arising from the stochastic encoding mechanism. Consequently, two storage processes may exhibit identical entropy while differing substantially in the organization and utility of their stored information. We therefore classify the information represented by $H(\Mn{t})$ into three complementary categories,
\begin{align}
    H(\Mn{t})
\Longrightarrow
    \left\{
        H_{\mathrm{use}}^{(n)},
        H_{\mathrm{red}}^{(n)},
        H_{\mathrm{noise}}^{(n)}
    \right\},
\end{align}
where $H_{\mathrm{use}}^{(n)}$ denotes information that is relevant for prediction, $H_{\mathrm{red}}^{(n)}$ denotes information that is duplicated or otherwise unnecessary within the information-storage state, and $H_{\mathrm{noise}}^{(n)}$ denotes variability arising solely from the stochasticity of the encoding mechanism.

This classification is conceptual and should not be interpreted as a mathematically rigorous partition of Shannon entropy. In general, useful information, redundant information, and encoder-induced variability do not constitute additive entropy components. Rather, the classification serves to distinguish the principal factors underlying $H(\Mn{t})$, whose operational definitions are introduced in the subsequent sections.

\subsection{Information Enthalpy Acquisition}
\label{sec:IEacquistion}

Information Enthalpy can be considered as a resource the system will use mainly for two purposes: 1. to predict future sensory inputs $O_{t+1}$, and facilitate a certain action (based on a policy $\pi$), and 2. to ``self-regulate'', that is, to change its own internal states $X_{t+1}$ to maintain a particular dynamics (such as critical dynamics).
To define the nature of this resource and map out how it is acquired and subsequently consumed, we must first look at how it builds within the system. For a meaningful pool of information enthalpy to accumulate, it must be actively acquired through the ongoing integration of both historical sensory inputs $O$ and system states $X$.

This update process highlights the open-system nature of the architecture. Because a closed, isolated system cannot spontaneously generate or maintain predictive structure, the maintenance of complex states like neural criticality depends entirely on a continuous influx of structured information. By actively embedding the multi-scale dependencies found within the joint history of $O$ and $X$ into the information-storage states $\big(\Mn{t}\big)$, the system prevents the natural decay of its predictive organization. To formalize these dynamics, we first introduce the \textit{information throughput capacity} from $P$ to $\Mn{t}$, that is, the rate at which this joint information is encoded into the system, and the predictive content of the information-storage at a given time-point $t$.

\subsubsection{Information throughput capacity.} 

To describe the mechanisms that lead the system to acquire information enthalpy, we start by defining the Information Throughput Capacity (ITC) from the past sensory inputs and system states to the information pool. This process is quantified by a directed-information rate~\cite{tsur_directed_2026} describing the causal information flow from the observed interaction history $P_{-\infty,t}$ to the storage state $\Mn{t}$, at each time-point $t$. 

\begin{figure}
    \centering
    \includegraphics[width=1.0\linewidth]{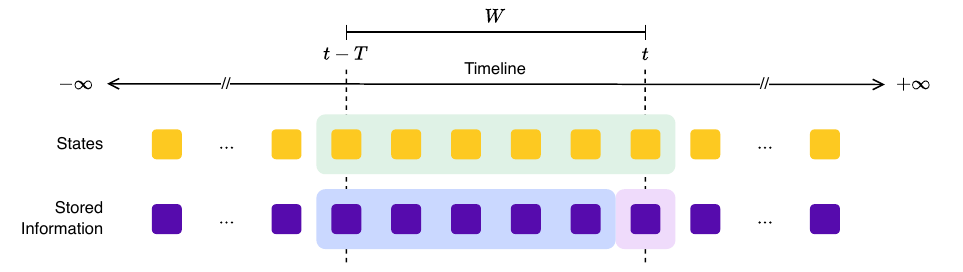}
    \caption{
    Schematic calculating directed information for a given observation window ($W$), as the relationship between the current information pool as a basic memory process $\tagMemoryCurrent{\Mn{t}}$ and sequence of past states $\tagStatesPast{P_{(t-T):t}}$, given the sequence of past information $\tagMemoryPast{\Mn{(t-T):(t-1)}}$.
    }
    \label{fig:through_capacity_schematic}
\end{figure}

To define the ITC, we first consider the directed-information~\cite{massey_causality_1990, marko_bidirectional_1973} over a linearly ordered temporal window $W_P$ of length $T$, ending at time-point $t$. We denote the onset of the observation period with $t_0 = \min(W_P) = t-T$, such that $W_P = \{t_0, t_0+1, \dots, t\}$. Thus, the window spans \(T\) unit-time intervals and contains \(T+1\) indexed states. The directed information from the past interaction history $P^T$ to the storage states $M^{(n)}_{t}$ over the window $W$ is defined as:
\begin{align}
    I\big(P^T\!\to\!M^{(n),T}\big)
=
    \sum_{j\in W_P}
    I \big(\tagStatesPast{P_{t_0:j}};\,\tagMemoryCurrent{\Mn{j}}\,\big|\,\tagMemoryPast{\Mn{t_0:j-1}}\big)
\label{eq:directed-info}
\end{align}
where $P_{t_0:j}$ and $\Mn{t_0:j-1} = (\Mn{t_0},\Mn{t_0 +1}, \dots, \Mn{j-1})$ are, respectively, the sequence of joint past states and the sequence of information-storage states, from the start of the window up to time-point $j-1$ (see~\Cref{fig:through_capacity_schematic}). $P^T$ and $M^{(n),T}$ are the vectors representing the entire sequence of joint history and information over the window $W_P$. The term \( I \big(P_{t_0:j};\,\Mn{j}\,\big|\,\Mn{t_0:j-1}\big)\) represents the ``innovation information'' at each time point $j$, quantifying the portion of the information flow to $\Mn{j}$ that is uniquely novel and not already present in the history of the information-storage states.

We then define the ITC at a timepoint $t$ as the average of the directed information~\eqref{eq:directed-info} over the finite temporal bound $T_{\mathrm{cap},t}$: 
\begin{align}
    \mathcal{R}^{(n)}_t
\coloneqq 
    \frac{1}{T_{\mathrm{cap},t}}
    \sum_{j \in W_{T_{\mathrm{cap}}}} 
        I \big(P_{t_0:j};\,\Mn{j}\,\big|\,\Mn{t_0:j-1}\big), 
\label{eq:DI-rate_finite}
\end{align}
where $t_0 = t - T_{\mathrm{cap}}$, and $W_{T_{\mathrm{cap}}} = \{ t_0, t_0 +1, \ldots, t\}$ is a linearly ordered window, of length ${T_{\mathrm{cap}}}$. Here, $\mathcal{R}_{t}^{(n)}$ is the local structural inflow rate of the specific process $\cMn$ evaluated at time step $t$. This definition is consistent with the definition of the long-term information rate as introduced in~\cite{tsur_directed_2026}, (see appendix~\ref{appsec:info-rate}).

Note that the directed-information rate from the joint history $P_{t_0, t}$ to each storage state $\Mn{t}$ is not unbounded, but it is strictly constrained by the system's throughput capacity $C_{\mathrm{r},t}\ge 0$ at each time-point $t$. Furthermore, admissible storage processes must also satisfy the storage-capacity constraint $H(\Mn{t}) \le \Cst{t}$ introduced in ~\eqref{eq:entropyMcap}.

Therefore, from the set of all possible storage processes $\mathfrak{M}$, we isolate the subset of admissible processes that satisfy this capacity budget across the system's entire operational lifespan. This restricted collection is defined as:
\begin{equation}
    \overline{\mathfrak{M}}(\mathbf{C}_{\mathrm{r}}, \mathbf{C}_{\mathrm{st}})
\coloneqq
    \Big\{ 
        \cMn \in \mathfrak{M} \, 
    \Big| 
        \mathcal{R}_{t}^{(n)} \le C_{\mathrm{r}, t}, \,
        H(\Mn{t}) \le \Cst{t}, \;
        \forall t \in \mathbb{Z} 
    \Big\},
\label{eq:admissible}
\end{equation}
This admissible set imposes two complementary resource constraints. The first is a flow constraint ensuring that the causal information inflow into the storage process at each time point cannot exceed the system’s storage throughput $\Cr{t}$. The second is a storage constraint requiring that the average information necessary to specify the realized value of each admissible information-storage state $\Mn{t}$ cannot exceed the actively maintainable storage capacity $\Cst{t}$.

\subsubsection{Scale Resolved Directed-Information Rate.}

While the baseline directed-information rate $R_t^{(n)}$ captures the information throughput capacity at the fundamental observation scale, interactions with the environment  typically present structured configurations spanning multiple concurrent time scales. To resolve these features, we extend this formulation to a discrete, scale-resolved directed-information throughput rate. This rate accounts for the joint past and information storage state trajectory across a downsampled history window, expressed as:
\begin{equation}
    \mathcal{R}^{(n),\tau}_{t} 
\coloneqq 
    \frac{1}{T_{\mathrm{cap},t}} \sum_{j \in W^\tau_{T_{\mathrm{cap}}}} 
    I\big(P^{\tau}_{t_0:j};\,\Mn{j}\,\big|\,\Mn{t_0:j-1}\big),
\label{eq:scale-resolved_DI}
\end{equation}
where $W^{\tau}_{T_{\mathrm{cap}}} $ is a macro-time window, constructed by stepping backward from the current time $t$ in uniform increments of scale $\tau$ such that every timestamp in the window is of the form $t - k\tau$. Here, $t_0 = \min (W^{\tau}_{T_{\mathrm{cap}}})$ is the earliest time-point on this window, within $T_{\mathrm{cap},t}$. This scale-resolved rate is obtained by applying a causal deterministic transform $\mathcal{G}^\tau$ to the interaction history $P$, which represents a scale-specific view of the past $P^{\tau}$ at a given timescale $\tau$. Therefore, the sequence $P^{\tau}_{t_0:j}$ is a downsampled history containing only the states that fall exactly on the macro-grid points from $t_0$ up to the current macro-index $j$. 
Hence, the sequence of past scale-specific information storage states $\Mntau{t_0:j-1}$ is evaluated strictly on the same macro-grid points.

Because structured information can be distributed across multiple concurrent scales, we operationalize an aggregate measure by combining these scale-resolved rates through a fixed discrete kernel with weights $w (\tau_j) \ge0$, normalized such that their sum equals 1:
\begin{align}
    \mathcal{R}^{(n)}_{\mathrm{st},t}
=
    \sum_{j = 1}^{j_{\textrm{max}}} 
        w(\tau_j) \, \mathcal{R}^{(n), \tau_j}_{t}
        .
\label{eq:ratemultiscale}
\end{align}
The time scales are ordered as $1 < \tau_1 \le \tau_2 \le \ldots \le \tau_{j_{\textrm{max}}} $. By restricting the aggregation exclusively to macroscopic time scales satisfying $\tau_j \ge \rho_j \tau_0$, where $\rho_j > 1$ and $\tau_0 = 1$, each individual component rate captures features that are temporally coarser than the raw data stream. Importantly, evaluating the information rate at a coarser temporal scale does not imply a lower rate value, since the conditional mutual-information contribution can itself change with the temporal scale at which the variables and their dependencies are evaluated.

The aggregate structured-information rate $\Rst{t}$ cannot exceed the system's throughput capacity:
\begin{equation}
    \mathcal{R}^{(n)}_{\mathrm{st},t} 
=
    \sum_{j = 1}^{j_{\textrm{max}}} w(\tau_j)\mathcal{R}^{(n),\tau_j}_t 
\le
    C_{\mathrm{r},t}
\end{equation}
Furthermore, the amount of information that can be extracted from any given signal is inherently bounded by its underlying statistical regularities, or structure. To account for this acquisition phase bound, we postulate that, at each time step, the structure available in the incoming signal (sensory input) can be characterized via the \textit{fractional availability of structured information} $\Gamma_t$.
This leads to the following constraint:
\begin{equation} 
    0 \,
\le\
    \,\mathcal{R}^{(n)}_{\mathrm{st},t} \,
    \le
\, \Gamma_t \, C_{\mathrm{r},t}.
\label{eq:structuredrateconstraint}
\end{equation}
More specifically, $0 \le \Gamma_t \le 1$ is a dimensionless, signal-dependent quantity characterizing the availability of structured information in the incoming signal over the set of temporal scales $\tau_j$ considered in the scale-resolved acquisition process. Thus, $\Gamma_t$ bounds the fraction of the instantaneous throughput capacity $\Cr{t}$ that can be occupied by multiscale structured information, rather than the amount necessarily acquired. This constraint is independent of the throughput constraint~\eqref{eq:DI-rate_finite} as the former bounds the multiscale components of the throughput, while the latter bounds the throughput at the finest scale.
In Sec.~\ref{sec:IEP}, we introduce the Information Enthalpy Potential (IEP) as a quantitative measure of the statistical structure of the input signal, which can be then used to characterize the structured-information availability represented by $\Gamma_t$.

We can therefore update the set of admissible information-storage processes $\overline{\mathfrak{M}}$, defined in Eq.\eqref{eq:admissible}, to include this third constraint, which bounds the aggregate scale-resolved rate $\mathcal{R}^{(n)}_{\mathrm{st},t}$ according to the amount of structured information available in the incoming signal
\begin{align}
    \overline{\mathfrak{M}}(\mathbf{C}_{\mathrm{r}}, \mathbf{C}_{\mathrm{st}}, \mathbf{\Gamma})
\coloneqq
    \Big\{ 
        \cMn \in \mathfrak{M} \, 
    \Big| 
        \mathcal{R}_{t}^{(n)} 
    \le 
        C_{\mathrm{r}, t}, \,
        H(\Mn{t}) 
    \le 
        \Cst{t}, \, 
        \Rst{t}
    \leq
        \Gamma_t \Cr{t},
    \;
    \forall t \in \mathbb{Z} 
    \Big\}
\end{align}
where we have introduced the trajectory $\mathbf{\Gamma} = \{\Gamma_t\}_{t \in \mathbb{Z}}$.

\subsection{Predictive content of information-storage pool and information enthalpy}

\paragraph{Predictive Content.} 
Once information has been stored in a state $\Mn{t}$, its utility is evaluated through its predictive content. For this purpose, we first consider the predictive information held in $\Mn{t}$ regarding a future time window $W_F = \{t+1, \dots, t+H\}$ of length $H$, referred to as \textit{horizon}. This can be defined via the conditional mutual information \(I(\Mn{t};F_{t+1:t+H}\mid A_{t:t+H-1
})\), where the conditioning on the future action plan $A_{t:t+H}$ filters out any components of the future that are predictable from the system's own intended actions.

Evaluating the information content at a single, arbitrary horizon provides only a partial snapshot of the total predictive information of $\Mn{t}$. Instead, we want to characterize the complete predictive profile across several horizons. To this purpose, we introduce a set of horizons \(\mathcal{K} = \{H_1, \dots, H_N \}\), with $H_1 < H_2 < \cdots <H_N$. As $\Mn{t}$ has a maximal past temporal reach $T_{\mathrm{cap},t}$, we bound the furthest time horizon as $H_N = T_{\mathrm{cap},t}$. We define the \textit{kernel-weighted} predictive-information content of $\Mn{t}$ as the discrete sum: 
\begin{align}
    \mathcal P\big(\Mn{t}\big)
\coloneqq 
    \sum^{N}_{i = 1}
        w(H_i)I\!\big(\Mn{t} \,;\, F_{t+1:t+H_i} \,\big|\,A_{t:t+H_i-1}\big)
\end{align}
The kernel $w(H_i)$ assigns a relative weight to the predictive information available at each horizon $H_i \in \mathcal{K}$. The condition  $w(H_i)\ge0$ ensures that no future horizon gets a negative weight. The normalization $\sum_{i = 1}^{N} w(H_i) = 1$ ensures that $\mathcal{P}(\Mn{t})$ represents a weighted average of predictive utility across all horizons, rather than an unscaled sum. 
To coherently link the action sequence \(A_{t:t+H}\) to future states, we define the action selection as $A_t \sim \pi^\star(\,\cdot \mid \Mn{t}, O_t\,)$, where  $\pi^\star$ denotes a policy restricted to the region of action space associated with high informational utility. This restriction reduces the dimensionality of the predictive-content calculation, while the mechanism by which the system selects or adapts policies to increase information enthalpy is left to future work.

We can then define the predictive utility of the system as a functional of the process trajectory $\cMn$ as 
\begin{equation}
    \mathcal{P}\big[\mathcal{M}^{(n)}\big]
= 
    \Big(
        \mathcal P\big(\Mn{t}\big)
    \Big)_{t \in \mathbb{Z}}
\label{eq:process_trajectory_functional}
\end{equation}
This functional $\mathcal{P}\big[\mathcal{M}^{(n)}\big]$ maps the information storage process $\mathcal{M}^{(n)}$ to a sequence of predictive-information values evolving over time, where each element $\mathcal P\big(\Mn{t}\big)$ measures the predictive-information content of the process at that specific step $t$.

\begin{figure}
    \centering
    \includegraphics[width=1.0\linewidth]{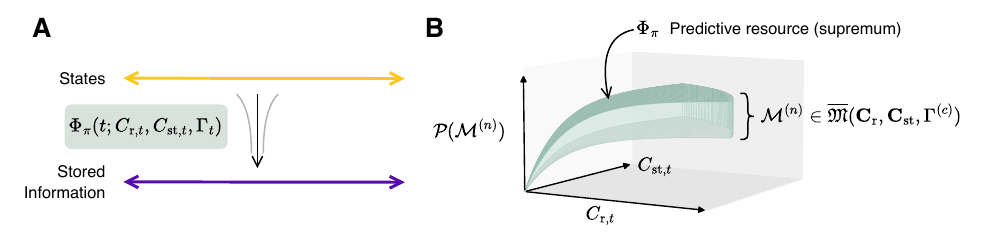}
    \caption{
    (\textbf{A}) Schematic indicating that information flow from states to stored information is rate limited by the predictive-resource hypersurface $\tagPredictiveResource{\PhiFun (t; \Cr{t}, \Cst{t}, \Gamma_t)}$ for a specific time-point $t$, and $\Gamma_t$.
    (\textbf{B}) The predictive-resource dynamics are non-decreasing and concave with respect to $C_{\mathrm{r},t}$, whereas the relationship with $C_{\mathrm{st},t}$ is not strictly concave, with illustrated concave relationship representing a special case. $\Gamma^{(c)}$ is kept constant for illustrative purposes.
    }
    \label{fig:predictive_resource_curve}
\end{figure}

\paragraph{Predictive-resource Hypersurface.}

To analyze how the system's storage constraints $\Rt{t} \leq \Cr{t}$ and $H(\Mn{t}) \leq \Cst{t}$ for a process $\cMn$, together with the signal-dependent structured-information constraint $\Rst{t} \le \Gamma_t \Cr{t}$, shape its predictive performance of an information-storage process $\cMn$ across time, we introduce the \textit{predictive-resource hypersurface}. Here we adopt a local comparative-static approximation. At each time \(t\), the various components of the constraint trajectories outside the variables displayed at \(t\) are treated as fixed background conditions. At each time step $t$, this is defined as the supremum of the predictive-information content $\mathcal{P}(\Mn{t})$
\begin{align}
    \tagPredictiveResource{\PhiFun (t; \Cr{t}, \Cst{t}, \Gamma_t)}
= 
    \sup_{\cMn \in \overline{\mathfrak{M}}(\mathbf{C}_{\mathrm{r}}, \mathbf{C}_\mathrm{st}, \mathbf{\Gamma})} 
    \, 
    \mathcal P\big(\Mn{t}\big).
\label{eq:PRC}
\end{align}
over the set of information-storage states $\Mn{t}$ whose trajectories $\cMn$ are in $\overline{\mathfrak{M}}(\mathbf{C}_{\mathrm{r}}, \mathbf{C}_\mathrm{st}, \mathbf{\Gamma})$. For the purpose of this formalization, the resulting conditional slice of the predictive-resource functional is written as \(\Phi_\pi(t ; \Cr{t},\Cst{t},\Gamma_t)\), suppressing its dependence on these fixed background conditions. This notation does not imply that predictive capacity is generally independent of the system’s history or future environment.

The supremum defines the upper predictive-information capacity of the system under the specified constraints; it does not require this bound to be attained by any particular admissible information-storage process. Thus, the predictive-resource hypersurface quantifies the predictive-information \textit{capacity} of the system at time $t$, for given $\Cr{t}$, $\Cst{t}$, and signal-dependent structured-information availability $\Gamma_t$.

We then define the \textit{predictive-resource hypersurface functional} as the mapping from the throughput- and storage-capacity trajectories $\mathbf{C}_{\mathrm{r}}$ and $\mathbf{C}_{\mathrm{st}}$, together with the trajectory $\mathbf{\Gamma}$, to the corresponding time-ordered sequence of predictive-information capacities:
\begin{equation}
    \PhiFun \big[\mathbf{C}_{\mathrm{r}}, \mathbf{C}_{\mathrm{st}}, \mathbf{\Gamma}\big]
\coloneqq
    \left\{
        \Phi_\pi (t;\Cr{t}, \Cst{t}, \Gamma_t)
    \right\}_{t \in \mathbb{Z}},
\label{eq:PRC_functional}
\end{equation}
The functional is therefore the time-ordered sequence of predictive-resource hypersurfaces defined pointwise by Eq.~\eqref{eq:PRC}. The supremum domain is the set of admissible processes $\overline{\mathfrak{M}}(\mathbf{C}_{\mathrm{r}}, \mathbf{C}_{\mathrm{st}},\boldsymbol{\Gamma})$ satisfying the constraints $\mathcal{R}_t^{(n)} \leq \Cr{t}$, $H(\Mn{t}) \leq \Cst{t}$, and $\Rst{t} \leq \Gamma_t\Cr{t}$ at every time step $t$ along the entire trajectory. We assume that $\PhiFun$ is non-decreasing in both capacity variables and concave with respect to $\Cr{t}$, such that increases in throughput capacity produce diminishing returns in predictive-information capacity. No analogous concavity assumption is imposed with respect to $\Cst{t}$, since storage capacity is a derived dynamical quantity and is not independent of the throughput capacity $\Cr{t}$ (see Sec.~\ref{sec:IEchanges}).

Importantly, the supremum defines the predictive-information capacity and does not specify the mechanism by which the system approaches or exploits this capacity; such a mechanism will depend on the dynamics of the considered system.

It is important to distinguish realized predictive information from predictive-information capacity. The quantity $\mathcal{P}(\Mn{t})$,  quantifies the \textit{realized predictive information} of the information-storage states $\Mn{t}$; that is, it describes the predictive information actually contained in a particular state $\Mn{t}$. By contrast, $\PhiFun(t;\Cr{t},\Cst{t}, \Gamma_t)$ quantifies the predictive-information capacity of the system. Consequently,
\begin{align}
    \mathcal{P}(\Mn{t}) \le \PhiFun (t;\Cr{t},\Cst{t}, \Gamma_t)
\end{align}
for every admissible process.

\paragraph{Information Enthalpy.}

Information Enthalpy can now be defined in terms of the predictive-resource hypersurface $\PhiFun (t;\Cr{t}, \Cst{t}, \Gamma_t)$ through a dimensional scaling factor:
\begin{equation}
    \HI(\Cr{t}, \Cst{t}, \Gamma_t) 
\coloneqq
    \kappa\ \PhiFun (t;\Cr{t}, \Cst{t}, \Gamma_t)
\label{eq:HI}
\end{equation}
Here, $\kappa$ is a conversion constant appropriate for the system's physical domain, intended to translate information bits of predictive information into a physical resource as stored within a neural system that enables an action through its action on the system's states. Here, \(\kappa\) is a dimensional scaling factor; it does not imply that predictive information is converted into energy or that information enthalpy constitutes a specific energy reservoir.  Two options are considered to define $\kappa$: a thermodynamic-based theoretical lower bound given by the Landauer limit per bit erased in a logically irreversible operation ($\kappa_\mathrm{td}{}=k_BT\ln 2$) or a biologically based factor ($\kappa_{\mathrm{bio}}$), which would have to be determined through empirical testing.

Analogously to the predictive-resource functional, we define the information enthalpy functional as
\begin{equation}
    \mathcal{H}_I
    \big[
        \mathbf{C}_{\mathrm{r}} , \mathbf{C}_\mathrm{st}, \mathbf{\Gamma}
    \big] 
\coloneqq
    \left\{
        \HI(t;\Cr{t}, \Cst{t}, \Gamma_t)
    \right\}_{t \in \mathbb{Z}}.
\label{eq:enthalpy_functional}
\end{equation}

A corresponding distinction can be made between the realized predictive-information resource and information enthalpy. For a particular information-storage state $\Mn{t}$, we define the realized predictive-information resource as $h_{I}(\Mn{t}) = \kappa \mathcal{P}(\Mn{t})$, with the corresponding process-level functional $h_I[\cMn] = \kappa \mathcal{P}[\cMn]$. These quantities represent predictive-information resource actually realized by a particular state $\Mn{t}$, or equivalently process $\cMn$. Information enthalpy, by contrast, represents the predictive-information resource capacity of the system under the specified constraints, and need not be attained at a given time step. Therefore:
\begin{align}
    h_{I}(\Mn{t}) 
\le 
    \HI(t;\Cr{t}, \Cst{t}, \Gamma_t). 
\end{align}

\subsection{Biological Constraints on Information Enthalpy}
\label{sec:biocosts}

Within a system characterized by an information enthalpy capacity $\HI$, the realized predictive resource $h_I(\Mn{t})$, associated with the predictive information contained in $\Mn{t}$, can be used to support interactions with the environment and self-regulation. 
Under an optimal policy \(\pi\), this predictive information can support two distinct functional pathways:
\begin{itemize}
    \item \textbf{Prediction of future sensory inputs:} Anticipate the next sensory observation $O_{t+1}$, facilitating effective interaction with the environment through actions $A_t \in \mathbb{R}^p$ selected according to policy $\pi$.
    \item \textbf{Internal Self-Regulation:} Regulate future internal neural states $X_{t+1}$, enabling the system to maintain critical dynamics and homeostatic stability. 
\end{itemize}
The use of this information does not imply its consumption or removal from the information-storage state. However, its acquisition, maintenance and use rely on biological processes subject to energetic and physiological constraints.

More generally, while an idealized information-processing system could be considered independently of the physical costs required for its operation and maintenance, real biological systems are necessarily subject to such costs. The acquisition, maintenance and use of realized predictive resource, together with the maintenance and adaptation of information enthalpy, depend on neural activity, plasticity, action implementation, and homeostatic regulation, all of which require biological resources. These requirements constrain the biologically attainable information-storage processes and capacity trajectories and, consequently, the realized predictive-information resource $h_I$, the predictive-resource hypersurface $\Phi_\pi$, and information enthalpy $\HI$.

To account for these constraints, we consider the total energy available for metabolism $E_t^{\mathrm{sup}}$. This finite energy budget supports both metabolic processes required to sustain the system, and its predictive information operations.

We distinguish three principal metabolic energy uses. The neural activity energy $E_t^{\mathrm{neural}}$ represents the energy required to implement the system’s neural dynamics, including modification of the internal states $X_{t}$, storage of incoming information at rate $\Rst{t}$, internal homeostatic processes, and neural processing associated with the use of stored predictive information. The plasticity energy $E_t^{\mathrm{plastic}}$ represents the energy required to modify the throughput and storage capacities $\Cr{t}$ and $\Cst{t}$. The action energy $E_t^{\mathrm{act}}$ represents the metabolic cost of actions through which the system interacts with the environment. The subscript $t$ indicates that each of these energies may vary over time. 

Neural systems must operate within a bounded homeostatic range to prevent structural damage, such as neural tissue damage from prolonged hyperexcitation. System dynamics that exceed these setpoints beyond viable limits, incur a penalty $\Psi(h_t)$. Here, $h_t$ denotes the homeostatic deviation from the tolerance radius $h_{\mathrm{tol}}$, at time $t$; for values of $h_t$ below $h_\mathrm{tol}$ the penalty remains negligible\footnote{An example of penalty function $\Psi(h_t)$ is provided in Appendix~\ref{appsec:homeostaticpenalty}}. Homeostatic requirements therefore provide an additional energetic constraint on viable system trajectories. We associate this penalty with an energetic cost $E^\mathrm{homeo}_t = \rho_h (\Psi(h_t))$, where $\rho_h$ maps the homeostatic penalty to the corresponding energetic cost.

The total metabolic energy required to sustain system operations is then:
\begin{equation}
    E_t^{\mathrm{req}} 
=   
    E_t^{\mathrm{neural}} + E_t^{\mathrm{plastic}} + E_t^{\mathrm{act}} + E^\mathrm{homeo}_t,    
\end{equation}
and energetic feasibility requires the metabolic-energy quantities $E_t^{\mathrm{req}} \le E_t^{\mathrm{sup}}$ which are distinct to the informational resources described elsewhere.  

The finite metabolic supply therefore imposes an additional energetic constraint on the information-storage processes that can be biologically realized. Neural energetic costs constrain implementation of the storage process; plasticity costs constrain attainable changes in $\Cr{t}$ and $\Cst{t}$; action costs constrain energetically feasible actions and policies; and homeostatic requirements restrict the viable state trajectories. These biological constraints do not constitute direct losses of information enthalpy, but further restrict the set of information-storage processes, capacity trajectories and actions that are biologically attainable. The supremum defining $\Phi_\pi$, and hence $\HI$, is therefore taken over the domain of processes satisfying both informational and biological feasibility constraints

We therefore update the set of admissible processes $\overline{\mathfrak{M}}(\mathbf{C}_{\mathrm r},\mathbf{C}_{\mathrm{st}},\boldsymbol{\Gamma})$ to include biological energetic constraint:
\begin{align}
    \overline{\mathfrak{M}}_\mathrm{bio}(\mathbf{C}_{\mathrm r},\mathbf{C}_{\mathrm{st}},\boldsymbol{\Gamma}, \mathbf{E}^\mathrm{sup}_t)
\coloneqq
    \Big\{ 
        \cMn \in \overline{\mathfrak{M}}(\mathbf{C}_{\mathrm r},\mathbf{C}_{\mathrm{st}},\boldsymbol{\Gamma}) \, 
    \Big| \,
        E^{\mathrm{req}}_t
    \leq
        E^{\mathrm{sup}}_t,\;
        \forall t \in \mathbb{Z} 
    \Big\},
\end{align}
where we have defined the trajectory $\mathbf{E}^\mathrm{sup}_t = \{E^\mathrm{sup}_t\}_{t \in \mathbb{Z}}$. Passive degradation of the information-processing capacities is represented by the decay terms in~\eqref{eq:maxstorecapacity} and~\eqref{eq:storagerate}, while degradation of predictive information already stored by the system is treated separately in Sec.~\ref{sec:IEchanges}.

The definitions of predictive hypersurface $\Phi_\pi$ and information enthalpy can be updated to include these energetic constraints as well:
\begin{align}
    \Phi_\pi (t; \Cr{t},\Cst{t}, \Gamma_t, E^\mathrm{sup}_t)
= 
    \sup_{\cMn \in \overline{\mathfrak{M}}_\mathrm{bio}}
    \mathcal{P}(\Mn{t}),
\end{align}
and
\begin{align}
    \HI (t; \Cr{t},\Cst{t}, \Gamma_t, E^\mathrm{sup}_t) 
=
    \kappa \Phi_\pi (t;\Cr{t},\Cst{t}, \Gamma_t, E^\mathrm{sup}_t).
\end{align}

\subsection{Changes in Information Enthalpy} 
\label{sec:IEchanges}

Having defined information enthalpy, we now consider how it may change over time. Such changes may reflect the changes in the capacities that constrain information acquisition and retention ($\Cr{t}$ and $\Cst{t}$), changes in structured information available in the incoming signal ($\Gamma_t$), and changes in the total energy available for biological processes ($E^\mathrm{sup}_t$). We collect these constraints into a generalized constraint vector
\begin{equation}
    Q_t
\coloneqq
    \left(
        \Cr{t},
        \Cst{t},
        \Gamma_t,
        E^{\mathrm{sup}}_t
    \right),
\end{equation}
which defines a point in the corresponding constraint space $\mathcal{Q}$. The predictive-resource hypersurface and information enthalpy can therefore be written compactly as
\begin{equation}
    \Phi_{\pi,t}=\Phi_\pi(t; Q_t),
    \qquad
    \mathcal{H}_{I,t}=\kappa\Phi_\pi(t;Q_t).
\end{equation}
The explicit dependence on $t$ allows the predictive-resource hypersurface $\Phi_{\pi,t}$ to vary even when the constraints $Q_t$ are held fixed. This accounts for changes in the joint statistics of  past states, encoded in $\Mn{t}$), and future states $F$, under which predictive information is evaluated. Thus, changes in $\HI$ may arise either from changes in the constraints defining the admissible predictive-resource capacity or from changes in the underlying statistical context. In the following, we first characterize changes associated with the individual constraints while holding the explicit temporal dependence fixed.

Changes in constraints are generally not independent: for instance, increasing $\Cr{t}$ may provide little additional predictive capacity if $\Gamma_t$ is very small. Furthermore, the variables themselves can be dynamically coupled. For instance, changes in $\Cr{t}$ and $\Cst{t}$ are associated with energy costs $E^\mathrm{plast}_t$, which are bound by $E^\mathrm{sup}_t$. Therefore, finite changes in $\Phi_\pi$ associated with a simultaneous change in several constraints cannot, in general, be identified with the sum of the independently evaluated single-constraint changes without specifying a path through $\mathcal Q$. 

Therefore, to characterize their respective effects, we consider changes in $\Phi_\pi(t;Q_t)$ with respect to each constraint separately, while holding the remaining constraints and the explicit temporal argument $t$ fixed. For each constraint $Q_{t,i}$, we define the constraint-specific finite change
\begin{equation}
    \Delta_i \Phi_{\pi,t}
    \coloneqq
    \left.
    \Delta \Phi_\pi(t;Q_t)
    \right|_{\{\Delta Q_{t,j}=0,\;j\neq i\}},
\end{equation}
and its corresponding finite-difference conjugate variable
\begin{equation}
    Y_{i,t}
    \coloneqq
    \frac{\Delta_i\Phi_{\pi,t}}
         {\Delta Q_{t,i}},
    \qquad
    \Delta Q_{t,i}\neq 0.
\end{equation}
The constraint-specific change in the predictive-resource hypersurface can therefore be written as
\begin{equation}
    \Delta_i\Phi_{\pi,t}
    =
    Y_{i,t}\Delta Q_{t,i}.
\end{equation}
Since $\mathcal{H}_{I,t}=\kappa\Phi_{\pi,t}$, the corresponding change in
information enthalpy is
\begin{equation}
    \Delta_i\mathcal{H}_{I,t}
    =
    \kappa Y_{i,t}\Delta Q_{t,i}.
\end{equation}
The variable $Y_{i,t}$ therefore quantifies the finite marginal change in $\Phi_\pi$ per unit change in the corresponding constraint, with all other constraints held fixed. The relation $Y_{i,t}\Delta Q_{t,i}$ has the mathematical form of generalized work, in which a change in a generalized coordinate is weighted by its conjugate generalized force~\cite{arnold_mathematical_1989}. Motivated by this structural analogy, we loosely interpret the variables $Y_{i,t}$ as \textit{conjugate predictive-resource forces}. This terminology denotes their conjugate role within the predictive-resource hypersurface, and does not imply that $Y_{i,t}$ constitute thermodynamic forces. Because the constraints defining $\Phi_\pi$ may be coupled, independently evaluated constraint-specific changes cannot in general be summed to obtain the change associated with a simultaneous variation of several constraints without specifying a path through  $\mathcal Q$. Furthermore, although multiplication by $\kappa$ expresses these changes in information-enthalpy units, the present framework does not establish that changes in information enthalpy are equivalent to thermodynamic work.

\paragraph{Capacity-induced Changes.} 

We first consider changes in $\Phi_\pi$ arising from variations in $\Cr{t}$ and $\Cst{t}$ while holding incoming-signal and energy changes fixed, and assuming the capacity changes to be well within the signal and energy constraints. Because $\Cr{t}$ and $\Cst{t}$ are dynamically related through Eqs.~\eqref{eq:maxstorecapacity} and~\eqref{eq:storagerate}, we consider their joint capacity-induced change using a specified ordering\footnote{For notational clarity, we suppress the fixed arguments $\Gamma_t$ and $E_t^{\mathrm{sup}}$ in what follows.}
\begin{align}
    \Delta_{\mathrm{cap}}\Phi_{\pi,t}
&  
=
    \Phi_\pi(t;\Cr{t}, \Cst{t})
-
    \Phi_\pi(t;\Cr{t-1},\Cst{t-1})
    \nonumber
    \\
&
=
    \Phi_\pi(t;\Cr{t}, \Cst{t})
-
    \Phi_\pi(t;\Cr{t}, \Cst{t-1})
+
    \Phi_\pi(t;\Cr{t}, \Cst{t-1})
-
    \Phi_\pi(t;\Cr{t-1},\Cst{t-1})
    .
\end{align}
We therefore define the localized \textit{marginal throughput utility} and \textit{marginal storage utility}, respectively, as   
\begin{align}
    \beta_t 
&
\coloneqq
    Y_{C_r,t}
=
    \frac{\Delta_\mathrm{cap} 
        \Phi_\pi 
        \left(t;
            \Cr{t}, \Cst{t-1}
        \right)}
        {\Delta \Cr{t}}
    ,
    \qquad
    \zeta_t 
\coloneqq 
    Y_{C_{st},t}
=
    \frac{\Delta_\mathrm{cap} \Phi_\pi (t;\Cr{t}, \Cst{t})}{\Delta \Cst{t}}.
\end{align}
Thus, $\beta_t$ measures the finite marginal change in predictive-information capacity per unit change in throughput capacity, whereas $\zeta_t$ measures the corresponding marginal change per unit change in storage capacity. They have the mathematical form of generalized forces conjugate to $\Cr{t}$ and $\Cst{t}$, respectively, along the specified path in capacity space.

Because $\Phi_\pi$ is assumed to be concave with respect to $\Cr{t}$, $\beta_t$ is non-increasing as throughput capacity increases, capturing diminishing returns from further increases in the rate at which predictive-resource capacity can be increased. By analogy with the conjugate relation between pressure and volume in thermodynamics, we refer to $\beta_t$ as an \textit{informational pressure} associated with throughput capacity.

Unlike $\beta_t$, diminishing returns are not generally assumed for $\zeta_t$, since no concavity condition is imposed on $\Phi_\pi$ with respect to $\Cst{t}$. A large positive value indicates that increasing the available storage provides a substantial predictive benefit, whereas a small or vanishing value indicates that additional storage capacity provides little increase in the predictive resource.  

As we have defined a specific path order for changes in $\Cr{t}$ and $\Cst{t}$, the capacity-induced changes of $\Phi_\pi$ admit the decomposition $\Delta_{\mathrm{cap}}\Phi_{\pi,t} = \beta_t\Delta C_{\mathrm r,t}+\zeta_t\Delta C_{\mathrm{st},t}$ which separates the contributions arising from changes in throughput capacity and storage capacity along the realized trajectory in capacity space. This decomposition evaluates the throughput-capacity increment before the storage-capacity increment. When the predictive-resource hypersurface contains interactions between the two capacity variables, reversing this order produces different values of $\beta_t$ and $\zeta_t$, although the sum of their corresponding capacity-weighted contributions remains equal to the total capacity-induced change. This reflects the interaction, or non-separability,  of $\Phi_\pi$ with respect to  $\Cst{t}$ on $\Cr{t}$.

\paragraph{Changes due to structured information.} 

We now  consider changes in predictive information capacity (and information enthalpy) arising from variations in the fractional availability of structured information $\Gamma_t$ available in the incoming signal, while holding the system’s information-processing capacities and energetic constraints fixed. Following the general definition above, we define the conjugate predictive-resource force associated with $\Gamma_t$ as
\begin{align}
    \xi_t
\coloneqq
    Y_{\Gamma,t}
=
    \frac{\Delta_\Gamma\Phi_{\pi,t}}{\Delta\Gamma_t}.
\end{align}
The corresponding constraint-specific changes in predictive-information capacity and information enthalpy are therefore
\begin{equation}
    \Delta_{\Gamma}\Phi_{\pi,t}
=
    \xi_t\Delta\Gamma_t,
    \qquad
    \Delta_{\Gamma}\mathcal{H}_{I,t}
=
    \kappa\xi_t\Delta\Gamma_t.
\end{equation}
The value of $\xi_t$ may depend on the current information-processing capacities and energetic constraints, reflecting the coupling between the different constraints defining the admissible set.

In the present framework, we do not specify an explicit functional dependence of $\Phi_\pi$ on $\Gamma_t$, nor a dynamical law governing changes in $\Gamma_t$. The IEP introduced in Sec.~\ref{sec:IEP} is conceptually related to $\Gamma_t$, as both characterize structured information available in an input signal. A future extension could introduce a time-resolved IEP, evaluated over finite temporal windows, to characterize variations in $\Gamma_t$. The dependence of $\Phi_\pi$ on $\Gamma_t$ could be determined empirically or modeled using established approaches to predictive structure and resource-constrained prediction, including computational mechanics~\cite{crutchfield_inferring_1989}, predictive information~\cite{bialek_predictive_1999}, predictive rate-distortion~\cite{still_structure_2008, marzen_predicive_2016}, information-bottleneck methods~\cite{tishby_information_1999}, or measures of information dynamics such as active information storage~\cite{lizier_local_2012}.

\paragraph{Energy-budget Changes.} 
Finally, we consider changes in predictive-resource hypersurface arising from variations in the energetic constraint $E^\mathrm{sup}_t$. Following the general definition above, we define the conjugate predictive-resource force associated with $E^\mathrm{sup}_t$ as
\begin{equation}
    \chi_t
    \coloneqq
    Y_{E^\mathrm{sup},t}
    =
    \frac{\Delta_E\Phi_{\pi,t}}
         {\Delta E^\mathrm{sup}_t}
    .
\end{equation}
The corresponding changes in predictive-information capacity and information enthalpy are therefore
\begin{equation}
    \Delta_E\Phi_{\pi,t}
    =
    \chi_t\Delta E^\mathrm{sup}_t,
    \qquad
    \Delta_E\mathcal{H}_{I,t}
    =
    \kappa\chi_t\Delta E^\mathrm{sup}_t.
\end{equation}
In this current work, no particular functional dependence of $\Phi_\pi$ on $E^{\mathrm{sup}}_t$ is assumed, as this will depend on the characteristic metabolic properties of the considered system. However, already at this stage a more elaborate formulation could additionally account for energetic efficiency, for example by favoring lower $E^{\mathrm{req}}_t$ among admissible processes that approach the same predictive-information supremum, or by allowing actions to influence $E^{\mathrm{sup}}_t$. 

\paragraph{Total changes.}
The constraint-specific changes considered above describe variations in $\Phi_{\pi,t}$ associated with changes in $Q_t$ at fixed explicit time. More generally, $\Phi_{\pi,t}$ may also change at fixed constraints due to changes in joint distribution of $\Mn{t}$ and $F$, represented by its explicit dependence on time. Adopting an ordering in which this explicit temporal change is evaluated first, the total finite change can be decomposed as
\begin{equation}
    \Delta
    \Phi_{\pi,t}
=
    \underbrace{\Phi_\pi (t; Q_{t-1})
-
    \Phi_\pi (t-1; Q_{t-1})}_{\Delta_t \Phi_\pi}
+ 
    \underbrace{\Phi_\pi (t; Q_t)
- 
    \Phi_\pi (t; Q_{t-1})}_{\Delta_Q \Phi_\pi}
    .
\end{equation}
The term $\Delta_t \Phi_\pi$ captures changes associated with the evolving joint statistics of past and future states at fixed constraints, whereas $\Delta_Q \Phi_\pi$ captures arising from variations in the constraints for fixed past-future joint statistics.

\subsection[Changes in predictive-information resource]{Changes in predictive-information resource $h_I$}

Changes in predictive-information resource $h_{I}(\Mn{t})$ are defined in terms of the change of predictive-information content \(\mathcal{P}\) between successive points in time:
\begin{equation}
    \Delta 
    h_{I}(\Mn{t})
= 
    \Delta
    h^{(n)}_{I,t}
=
   \kappa 
    \left[
        \mathcal{P}(\Mn{t})
    -
        \mathcal{P}(\Mn{t-1})
   \right].
\label{eq:hIchangesgeneric}
\end{equation}
To provide a mechanistic interpretation of these changes, we propose that they can be described in terms of predictive-information inflow and outflow rates, $\mathcal{R}_\mathrm{pred,t}$ and $\mathcal{R}_\mathrm{out,t}$ respectively. We also propose that these rates can be defined in terms of specific following phenomenological contributions associated with the acquisition and loss of predictive information.

First, we considered the simplified assumption of linear proportionality between the predictive-information inflow rate $\mathcal{R}_\mathrm{pred,t}$ and the multiscaled structured information rate \(\Rst{t}\):
\begin{equation}
    \mathcal{R}_{\mathrm{pred},t}
\approx
    \eta_t \Rst{t}\, ,
\qquad
    0\leq\eta_t\leq1.
\end{equation}
The parameter $0 \le \eta_t \le 1$ represents the fraction of acquired structured information that contributes to the predictive information realized in $\Mn{t}$. Its value varies over time according to the system's intrinsic mechanisms of information conversion and also the metabolic resources available to support this process. For instance, insufficient energy resources ($E^\mathrm{sup}_t < E^\mathrm{req}_{\mathrm{conv},t}$) may reduce the ability of the system to convert acquired information into predictive information, lowering $\eta_t$. While this simplification is useful to illustrate the relation between $\mathcal{R}_{\mathrm{pred},t}$ and $\Rst{t}$, and thus changes in $h_I$, this dependence may in general be non-linear.

Similarly, realized predictive information may be lost through several mechanisms. We consider four principal contributions. Passive loss, $\mathcal{R}_{\mathrm{pass},t}$, accounts for intrinsic degradation of stored predictive information through processes such as forgetting, noise, or biological turnover. Use-related loss, $\mathcal{R}_{\mathrm{use},t}$, accounts for the fraction of predictive information that is destroyed, overwritten, or otherwise rendered unavailable as a consequence of its use. Importantly, the use of predictive information does not necessarily imply its loss; this term represents only the component that is actually lost through use. Maintenance loss, $\mathcal{R}_{\mathrm{maint},t}$, accounts for additional loss resulting from insufficient metabolic resources to maintain already realized predictive information, for instance when $E^{\mathrm{sup}}_t<E^{\mathrm{req}}_{\mathrm{maint},t}$. Finally, contextual loss, $\mathcal{R}_{\mathrm{context},t}$, accounts for retained information that ceases to be predictive because the statistical structure of the environment, or of the relevant future, changes, even when the stored information itself has not degraded. We therefore define the total predictive-information outflow rate as
\begin{equation}
    \mathcal{R}_\mathrm{out,t}
=
    \mathcal{R}_\mathrm{pass,t}
+
    \mathcal{R}_\mathrm{use,t}
+
    \mathcal{R}_\mathrm{maint,t}
+
   \mathcal{R}_\mathrm{context,t}
\end{equation}
Combining these contributions, we propose the following phenomenological balance law for changes in the realized predictive-information resource\footnote{As in previous definitions, the factor $\Delta t$ is omitted.}:
\begin{equation}
    \boxed{
    \Delta h^{(n)}_{I,t}
=
    \kappa
    \left[
        \underbrace{\eta_t \Rst{t}}_{\mathrm{inflow}}
    -
        \underbrace{(\mathcal{R}_\mathrm{pass,t}
+
    \mathcal{R}_\mathrm{use,t}
+
    \mathcal{R}_\mathrm{maint,t}
+
   \mathcal{R}_\mathrm{context,t}}_{\mathrm{outflow}})
    \right]}.
    \label{eq:hIbalance}
\end{equation}
Equation~\eqref{eq:hIbalance} is a phenomenological decomposition of the change defined in Eq.~\eqref{eq:hIchangesgeneric}, rather than an independent definition of $h_I$. Consequently, the net contribution of the proposed inflow and outflow mechanisms must recover the corresponding change in predictive-information content $\mathcal{P}$.


\section{Information Enthalpy Potential}
\label{sec:IEP}

Within any signal is a degree of structure, complexity or sophistication - depending on your chosen term. From a completely disorganized or random signal to one highly structured. It has been proposed before that the brain is tuned to respond to certain qualities of a signal based on the structure (which can be interpreted as information) it contains\cite{schmidhuber_formal_2010, schmidhuber_simple_2007, aaronson_quantifying_2014}. Quantifying this structure can be difficult as such structure can exist across multiple scales and a variety of dimensions. Yet without this quantification, it is unrealistic to be able to test how different types of structure might influence neural system behavior. 

To ensure that the ideas behind information enthalpy are testable, a concrete formalization and implementation of the Information Enthalpy Potential ($\mathrm{IEP}$) is now introduced. $\mathrm{IEP}$ is proposed as a dimensionless measure of the potential of environmental input to supply structured information that could contribute to the accumulation of predictive informational resources in a neural system that would be consistent with information enthalpy. While the quality of information enthalpy might be calculated by any number of measures, concretely we need some testable methods to quantify the extent to which this property is present. Such information structure should have key properties that are detectable, including limited compressibility, low entropy, and high statistical complexity across time. 

To quantify the extent to which a signal exhibits these properties, we define the $\IEP$ score as a function of two primary measures: the entropy-weighted redundancy score $R^{\alpha}(S)$, and the multiscale statistical complexity $C_{\MS}(S)$. Here, $R^{\alpha}(S)$ modifies the standard Lempel-Ziv complexity (LZC)  by incorporating an entropy weight (where $\alpha$ denotes  the entropy exponent) to evaluate the incompressibility of an input sequence $S$. This adjustment corrects the inherent tendency of LZC to overestimate stochastic noise.  $C_{\MS}(S)$ then generalizes this framework to capture informational structure across varying spatial and temporal scales.  The symbolic sequence $S $, provides a suitable representation of the signal. These components are integrated into the following general expression for the $\IEP$:
\begin{align} 
    \IEP (S ; \Theta)
=
    (1-\beta) R^{\alpha}(S) 
+ 
    \beta C_{\MS}(S)
\end{align}
where $\Theta$ represents the set of the hyperparameters. These encompass weighting factors, such as $\alpha$ and $\beta$, which modulate the relative influence of entropy and structural redundancy, and the encoding parameters that pertain to the translation of the raw signal into the symbolic sequence $S$. In particular, the parameter $\beta \in [0,1]$ acts as a weighting factor that balances the contribution of local redundancy against multiscale organizational complexity. By defining the $\IEP$ through these modified measures of incompressibility, we can effectively isolate genuine structural organization from stochastic fluctuations across a diverse hierarchy of scales. 

\paragraph{Sequence implementation.}

To derive the sequence $S$, a continuous signal is typically partitioned into \text{atomic symbols} or tokens~\footnote{We use the word ``atomic'' to denote fundamental, irreducible units of information within a specific framework.} through a discretization process tailored to the source's specific characteristics. This process ensures that the signal’s underlying dynamics are mapped onto a finite alphabet $\mathcal{A} = \left\{s_1, s_2, \ldots, s_M \right\}$ of $M$ unique atomic symbols. The signal is therefore represented as a symbolic sequence $\tagSymbols{S = {s_1 s_2 \ldots s_N}}$, where each element $s_i \in \mathcal{A}$, and $M \leq N$. The discretization is a critical step, as the choice of mapping rules determines which features of the signal are preserved and which are filtered out. 

Here, we briefly describe the process we implemented in this work to convert the raw signal into a sequence as illustrated in Fig~\ref{fig:symbolization}; 
the complete derivation is provided in Appendix~\ref{appsec:sequence_setup}. We start with a signal recorded from $C$ distinct channels over $T$ time-steps, collected in a 2D binary raster $\tagRaster{X\in \left\{0,1\right\}^{C \times T}}$. The raster $X$ is first partitioned into \textit{patches}, each of size $C_p \times T_p \in \bbN^2$, where the dimensions $C_p$ and $T_p$ are determined by the required ``scale of observation'' or ``coarse graining''. Notably, starting indices for each patch are selected using \textit{strides} $s_c \neq C_p $ and $s_t \neq T_p$. By setting $s_c < C_p, s_t < T_p$, the patches overlap, thus preserving spatial and temporal correlations across patches.
\begin{figure}[ht]
  \centering
  \includegraphics[width=\textwidth]{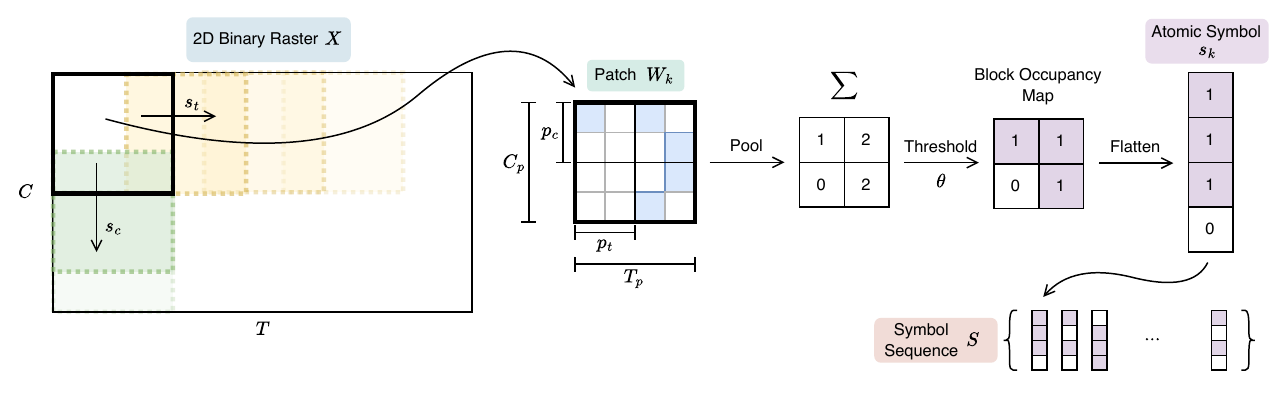}
  \caption{
  Schematic of deriving symbol sequence $\tagSymbols{S}$ from binary raster $\tagRaster{X}$ by taking patches $\tagPatch{W_k}$ and applying pooling, thresholding, and flattening within each patch to derive an atomic symbol $\tagSymbol{s_k}$.
  }
  \label{fig:symbolization}
\end{figure}
The transformation of a patch $\tagPatch{W_k}$ into an atomic symbol is done through a two-step pooling and flattening process. First, it is subdivided into non-overlapping blocks of size $p_c \times p_t$, where a block is assigned a value of 1 if its occupancy (defined as the fraction of active elements) exceeds a specified threshold value, and 0 otherwise. This coarse-graining operation effectively reduces the high-resolution patch to a pooled grid of dimensions $\lfloor C_p/p_c \rfloor \times \lfloor T_p/p_t \rfloor$. Subsequently, this pooled grid is flattened into a binary string that defines the atomic symbol $\tagSymbol{s_k}$.

\paragraph{Sequence Weighting.} To favor structured organization over stochastic noise, we transform $S$ into a weighted sequence $\tilde S$ by assigning to each symbol $s_k$ a weight $w_k$ derived using the Permutation Statistical Complexity (PSC), $\widehat C$.
By combining the permutation entropy $H_p$~\cite{bandt_permutation_2002} and a disequilibrium factor $Q(t)$ ~\cite{rosso_distinguishing_2007}, the PSC effectively distinguishes structured dynamics from purely random fluctuations. The weights are then used to define a replication count $\kappa_k \coloneqq 1 + \lfloor \lambda w_k\rfloor$, with $\lambda$ being the replication strength. This process yields a modified sequence $\tilde S$ consisting of concatenated repetitions of each atomic unit. This amplification accentuates organized regimes while preserving the original chronological order of the signal. Both the redundancy score $R^\alpha$ and multiscale complexity $C_{\MS}$ are evaluated for the modified sequence $\tilde S$.

\subsection{Entropy-Weighted Shuffled Lempel-Ziv Complexity and Redundancy Score}

The Lempel-Ziv complexity (LZC) metric has long been used as a method to evaluate the complexity (incompressibility) of a signal ~\cite{lempel_complexity_1976}. The measure has been used and adapted in a range of neuroscience-related explorations and, in specific contexts, is highly  predictive of key brain states \cite{hohn_spectral_2024, medel_complexity_2023}.  Yet, key limitations persist in separating random information (highly incompressible with standard approaches) against meaningfully complex signals~\cite{nagarajan_quantifying_2002, aboy_interpretation_2006, zozor_lempelziv_2005, bai_permutation_2015}. 

To address this issue, we derive a redundancy score using a modified version of the standard LZC. First, we calculate the standard LZC on the sequence $\tilde S$ to determine the number of phrases $c(\tilde S)$, required to capture the signal ($c(\cdot)$ is the LZC phrase count on a finite token sequence). We then randomly shuffle the sequence $\tilde S$ and re-calculate the complexity to obtain the expected phrase count under a random permutation $\sigma$. This value, $c_{\operatorname{shuf}}(\tilde S):=\E_{\sigma}\!\left[c(\sigma\cdot\tilde S)\right]$, is estimated using M surrogates. This procedure preserves the token histogram while destroying temporal dependence in a normalized fashion. The redundancy score is then
\begin{equation}
R(\tilde S) \; :=\; \frac{c_{\operatorname{shuf}}(\tilde S)-c(\tilde S)}{c_{\operatorname{shuf}}(\tilde S)} = 1- \frac{c(\tilde S)}{c_{\operatorname{shuf}}(\tilde S)} \;\in\;[0,1]_{+},
\label{eq:redudancy_score}
\end{equation}
with non-negativity clipping when the numerator is negative. As such, if $R \approx 1$ (where $c(\tilde S) \ll c_{\operatorname{shuf}}(\tilde S))$, indicates high temporal structure, while if  $R \approx 0$, that is $c(\tilde S) \approx  c_{\operatorname{shuf}}(\tilde S)$, there is no serial structure. 

The risk with this formulation is that it may reward trivial structure of extended runs of identical tokens. To penalize this type of structure we introduce the entropy-weighted redundancy score $R^{(\alpha)}(\tilde S)$, for an entropy exponent $\alpha\ge 0$, defined in terms of the \textit{normalized token entropy}
\begin{equation}
\Hsym(\tilde S)\;=\; \frac{-\sum_{a\in\mathcal{A}(\tilde S)} p(a)\,\log_2 p(a)}{\log_2 \lvert\mathcal{A}(\tilde S)\rvert}\ \in[0,1].
\end{equation}
Here, the probability distribution $p(a)$ is the empirical token distribution of the sequence $\tilde S$, and $\lvert\mathcal{A}(\tilde S)\rvert$ is the number of distinct tokens observed. Defined this way, $\Hsym(\tilde S)$ will be lower for sequences with trivial structure and higher for those exhibiting a wider variety of tokens. The entropy-weighted redundancy score of a given sequence is then:
\begin{equation}
R^{(\alpha)}(\tilde S) :=\ R(\tilde S)\,\big[ \Hsym(\tilde S) \big]^{\alpha} \in [0,1],
\end{equation} 
where $\alpha$ is the entropy exponent. This adjustment biases the redundancy score toward sequences with greater symbolic variety. For clarity, \(H_{\mathrm{sym}}=0\) when \(|\mathcal A|=1\), and \(R^{(0)}=R\).

\subsection{Multi-scale statistical complexity}

A key feature of information processed by neural systems is that it is presented and acquired over time at a given rate \cite{ay_information_2015, allen_multiscale_2017}. Therefore, it is necessary to assess whether any $\IEP$ present in a set of information persists throughout the duration of the signal. However, as the scale at which this structure exists may vary dramatically based on the information itself, a single scale is unlikely to capture this aspect suitably \cite{ay_information_2015, allen_multiscale_2017}. To capture this aspect of the statistical complexity across scales, we introduce a \textit{scale-conditional} redundancy score $R_B(\tilde S)$ defined in the following way. First, we consider a set of block sizes $\mathcal{B}=\{B_1<\cdots<B_m\}$, where each block size serves as a proxy for a distinct scale of structural organization. For each block size $B_i \in \mathcal{B}, i \in [1,m]$, we perform the following operations:
\begin{enumerate}
    \item Partition the sequence $\tilde{S}$ in contiguous blocks of length $B_i$ (in token units).
    \item Apply a block-permutation operator $\pi_B$ that permutes contiguous blocks while preserving order within each block; the resulting permuted sequence is denoted as $\tilde S_{B_i} = \pi_B \cdot \tilde S $. Notably, the functional form of $\pi_B$ is invariant to the choice of block size.
    \item Calculate the expected LZC phrase count for the resulting shuffled sequence: $c(\tilde S_{B_i}) = :=\mathbb E_{\pi_{B_i}}\left[c\!\left(\pi_{B_i}\tilde S\right)\right]$.
\end{enumerate}

While a small value of $B_i$ will disrupt local structure, a large $B_i$ will preserve the local structure yet scramble larger-scale order. For each block-size $B_i$ we define the scale-conditional redundancy score as : 
\begin{equation}
    R_B(\tilde S)\ :=\  1-\frac{c(\tilde S)}{c(\tilde S_{B_i}) },\qquad \forall \,B_i\in\mathcal{B}.
    \label{eq:multiscale_redundancy}
\end{equation}
This is a generalization for multiple scales, of the redundancy score defined in~\eqref{eq:redudancy_score}; specifically, for $B_1 = 1$, the measure in~\eqref{eq:multiscale_redundancy} reduces to the standard redundancy score $R(\tilde S) = R_1$, which is independent of of the multiscale set at $B$, provided that $\E\!\left[c(\pi_B\tilde S)\right]>0$.

We can then define a measure of the persistence of structure across scales by integrating~\eqref{eq:multiscale_redundancy}  in the following way. We first define the minimum value $x_{\min}=\min_B x_B$, $x_{\max}=\max_B x_B$, and $x_B=\log_2 B$

Let $x_{\min}=\min_B x_B$, $x_{\max}=\max_B x_B$, and $x_B=\log_2 B$ as block size will typically grow exponentially and so use of a logarithmic scale is likely to better capture long-range dependencies. Integrating across scales by calculating the normalized area under the curve is given as
\begin{equation}
C_{\MS}(\tilde S)\ :=\ \frac{\displaystyle \int_{x_{\min}}^{x_{\max}} R_B(\tilde S)\,\mathrm{d}x}{\displaystyle R_1(\tilde S)\,\big(x_{\max}-x_{\min}\big)}\ \in[0,1],
\end{equation}
and provides a measure of how the structure persists across scale. Trapezoidal integration over the sampled grid $\{x_B\}$ is chosen in this example as only discrete $R_B$ are available. However other methods may be more suitable in other scenarios.

\subsection{Information Enthalpy Potential}

Using the above, an $\IEP$ can be calculated as a composite score intended to quantify the maximum amount of structured information in a given signal that might be acquired by a system. Here, this is augmented by a mixing weight $\beta\in[0,1]$ to allow for different components to be emphasized, formalized as:
\begin{equation}
\boxed{\;\IEP(X;\Theta)\;:=\;(1-\beta)\,R^{(\alpha)}(\tilde S)\;+\;\beta\,C_{\MS}(\tilde S)\;\in[0,1]\;}.
\end{equation}
where $\Theta$ represents the hyperparameters selected for each component score. Explicitly these hyperparameters are:
\begin{equation}
\Theta := \big(
\underbrace{C_p,T_p,s_c,s_t,p_c,p_t,\theta}_{\Theta_{\mathrm{token}}},
\underbrace{\alpha,\beta,\mathcal{B},M}_{\Theta_{\mathrm{score}}},
\underbrace{D,\tau,w,\lambda,\nu}_{\Theta_{\mathrm{PSC}}}
\big),
\end{equation}
where $\nu$ denotes the PSC weight-normalization choice (e.g., min-max or mean) used in $w_k=\mathcal{N}(\bar w_k;\nu)$. 

\subsection*{Hyperparameter roles and interactions}
\begin{description}[leftmargin=2.2em,labelsep=0.6em]
\item[$C_p,T_p$ (patch sizes)]: Control spatial/temporal context of each token. Larger values coarsen the partition (fewer, longer-range tokens), changing both redundancy $R(\tilde S)$ and the multi-scale area by altering symbol statistics and token count.
\item[$s_c,s_t$ (strides)]: Set overlap between neighboring tokens. Increasing strides reduces overlap-induced repeats, typically decreasing spurious $R(\tilde S)$ on random data and sharpening discrimination.
\item[$p_c,p_t$ (pool sizes)]: Set coarse-graining prior to symbolization. Larger pools reduce alphabet richness and may increase repeats; this lowers $\Hsym$ and can inflate $R$, thus affecting $S_\alpha$.
\item[$\theta$ (occupancy threshold)]: Determines binary pooling. Higher $\theta$ sparsifies pooled bits (smaller effective alphabet), reducing $\Hsym$ and modulating $R$.
\item[$\alpha\ge 0$ (entropy exponent)] Scales the redundancy by token entropy: $\partial S_\alpha/\partial \alpha=(\ln \Hsym)\,R\,\Hsym^{\alpha}\le 0$ for $\Hsym\in(0,1]$. Thus larger $\alpha$ down-weights degenerate/low-entropy tokenizations.
\item[{$\beta \in [0,1]\; \text{(mixing weight)}$}]: 
Linear trade-off between local structure and scale-persistence: 
$\frac{\partial \IEP}{\partial \beta} = C_{\MS} - S_\alpha$. 
Larger $\beta$ emphasizes long-range organization captured by $C_{\MS}$.
\item[$\mathcal{B}=\{B_i\}$ (block scales)]: Chooses the block sizes for block-permutation surrogates. It sets the integration domain $[x_{\min},x_{\max}]$ with $x_B=\log_2 B$ and the sampling grid for the trapezoidal estimate; adding larger $B$ increases sensitivity to long-range dependencies.
\item[$M\in\bbN$ (surrogate count)]: Number of Monte-Carlo surrogates used for expectations in $c_{\mathrm{shuf}}$ and $\E[c(\pi_B\tilde S)]$. Larger $M$ reduces estimator variance but does not change the population definition of $\IEP$.
\item[$D,\tau,w$ (PSC parameters)]: Embedding dimension, ordinal delay, and window length for the PSC field $\widehat C_t$. They determine the statistics of $\bar w_k$ and hence token weights; larger $D$ or $w$ increases pattern resolution and stability (with higher sample demand).
\item[$\lambda\ge 0$ (PSC replication strength)]: Controls replication counts $\kappa_k=1+\lfloor\lambda w_k\rfloor$. Larger $\lambda$ increases the influence of high-PSC epochs on $\tilde S$, thereby modifying $R,\Hsym,C_{\MS}$ jointly. Setting $\lambda=0$ disables PSC weighting ($\tilde S\equiv S$).
\item[$\nu$ (PSC normalization)]: Choice of normalization $\mathcal{N}(\cdot;\nu)$ (e.g., min--max vs.\ mean) used to map $\bar w_k$ into $[0,1]$. It shapes the relative scale of $w_k$ and thus the replication distribution $\{\kappa_k\}$.
\end{description}
All interactions are mediated through $\tilde S=\tilde S(X;\Theta)$; in particular, any change to tokenization or PSC parameters alters the symbol sequence on which $R, \Hsym$ and the family $\{R_B\}_{B\in\mathcal{B}}$ (hence $C_{\MS}$) are computed, and the final $\IEP$ combines these effects through $\alpha$ and $\beta$.


\section{Assessment of IEP as a metric of Spatiotemporal Complexity}
\label{sec:IEP-dem}

Conceptually, $\IEP$ increases when a signal exhibits reproducible spatiotemporal organization across multiple scales and decreases for data that are either too regular or too random. We compute $\IEP$ on channel~$\times$~time rasters from two complementary datasets: 
(i) synthetic spike rasters generated from canonical fractal geometries, 
and (ii) biologically realistic spiking recordings from the publicly available \href{https://zenkelab.org/resources/spiking-heidelberg-datasets-shd/}{\textit{Spiking Speech Commands (SSC)} dataset \cite{cramer2020heidelberg}}, 
which provides event-based auditory data. 

For each original raster, we also created three surrogate controls designed to progressively disrupt structure while preserving different marginal statistics:
\begin{enumerate}[label=(\roman*)]
\item \textbf{Permute time}: independently permutes time within each channel (preserves per-channel spike counts; destroys temporal order).
\item \textbf{Bernoulli global}: samples an i.i.d.\ Bernoulli raster with the same global firing rate (preserves overall sparsity; removes channel/time dependencies).
\item \textbf{Permute all}: shuffles all entries and reshapes (preserves total spike count only).
\end{enumerate}
Hyperparameters used for all experiments are summarized in Table~\ref{tab:hyperparams_datasets}.

\subsection{Fractal-driven spike rasters from synthetic geometries}

\begin{figure}[!htp]
  \centering
  \includegraphics[width=\textwidth]{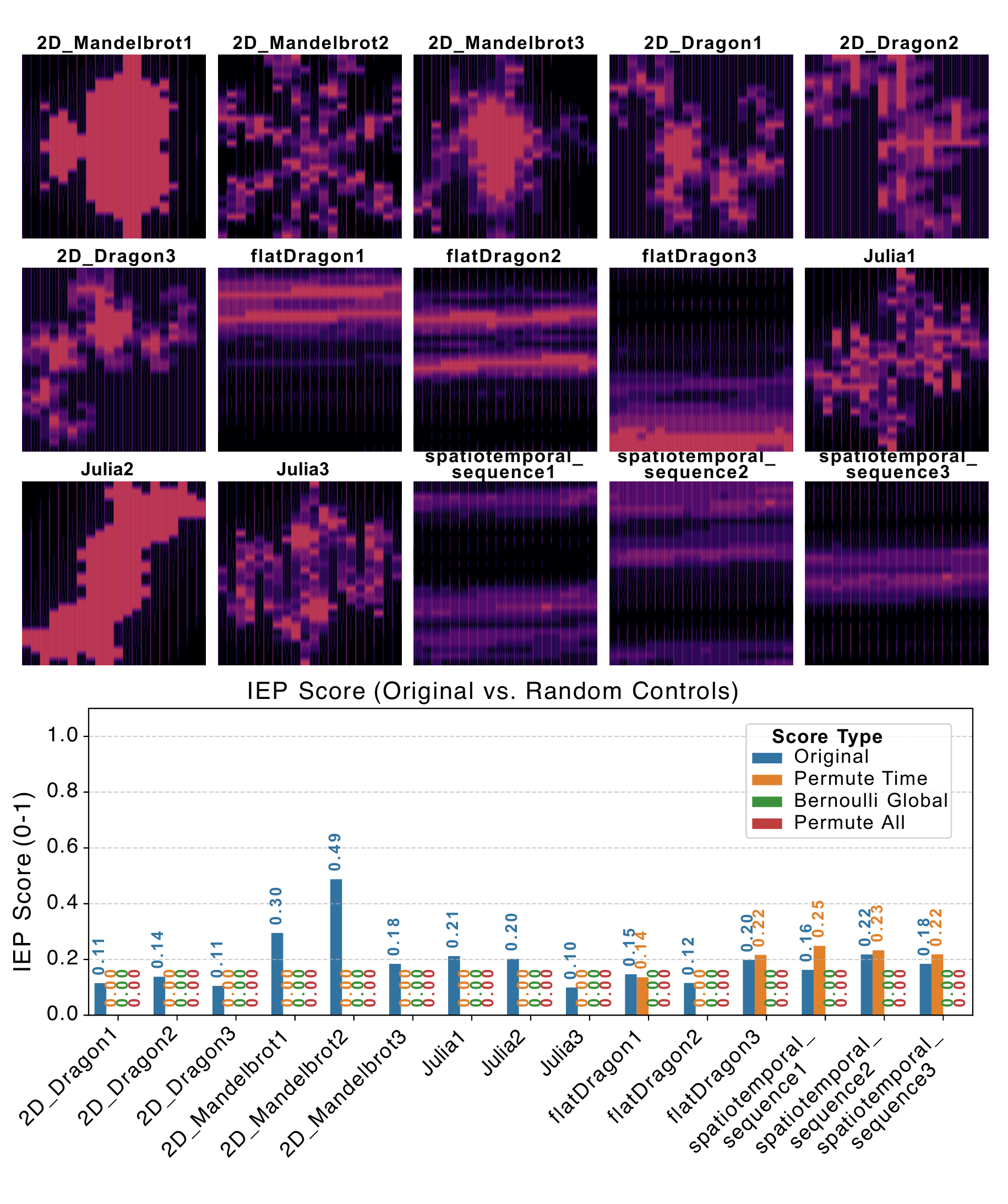}
  \caption{\textbf{$\IEP$ on fractal-derived spike rasters}. \emph{Top}: example binary channel--time rasters generated from canonical fractal geometries (Mandelbrot, Dragon, Julia) and three hand-crafted spatiotemporal sequences (interpolated shades indicative of spike density). \emph{Bottom}: grouped bars show $\IEP$ for the \textcolor{blue}{Original} patterns versus three surrogates: \textcolor{orange}{Permute Time} (per-channel temporal shuffling), \textcolor{green}{Bernoulli Global} (i.i.d.\ spikes at the same global rate), and \textcolor{red}{Permute All} (full shuffle). Originals tend to score higher, indicating sensitivity of $\IEP$ to multi-scale spatiotemporal structure rather than to simple firing-rate statistics. See Table~\ref{tab:hyperparams_datasets} for the dataset-specific hyperparameters.}
  \label{fig:iep_fractals}
\end{figure}

To test $\IEP$ on signals with \emph{a priori} structure that is rich yet non-periodic, we generated synthetic spike rasters from canonical fractal sets (e.g., Mandelbrot, Julia, and Dragon variants) as well as hand-crafted spatiotemporal sequences. Each pattern was rasterized onto a channel--time grid to yield binary spike maps with characteristic long-range structure and self-similarity (examples in the top panel of Fig.~\ref{fig:iep_fractals}). 

The bottom panel of Fig.~\ref{fig:iep_fractals} summarizes calculated $\IEP$ scores for all patterns and their controls. In most cases, the original fractal-derived rasters obtain higher $\IEP$ values than all surrogates, reflecting sensitivity of $\IEP$ to intrinsic, multi-scale structure. This separation is most pronounced for certain fractal sets such as the Mandelbrot and Julia patterns, whereas some of the manually designed spatiotemporal sequences yield comparable values between originals and controls due to their simpler or more homogeneous structure. Overall, these results illustrate that $\IEP$ captures the degree of intrinsic organization within the data rather than mere firing-rate statistics. These results support the intended behavior of $\IEP$: genuine structured signals (blue) score above randomized counterparts (orange/green/red) that conserve only simple marginals.

\subsection{Auditory-driven spike rasters from the Spiking Speech Commands (SSC) dataset}
\begin{figure}[!htp]
  \centering
  \includegraphics[width=0.93\textwidth]{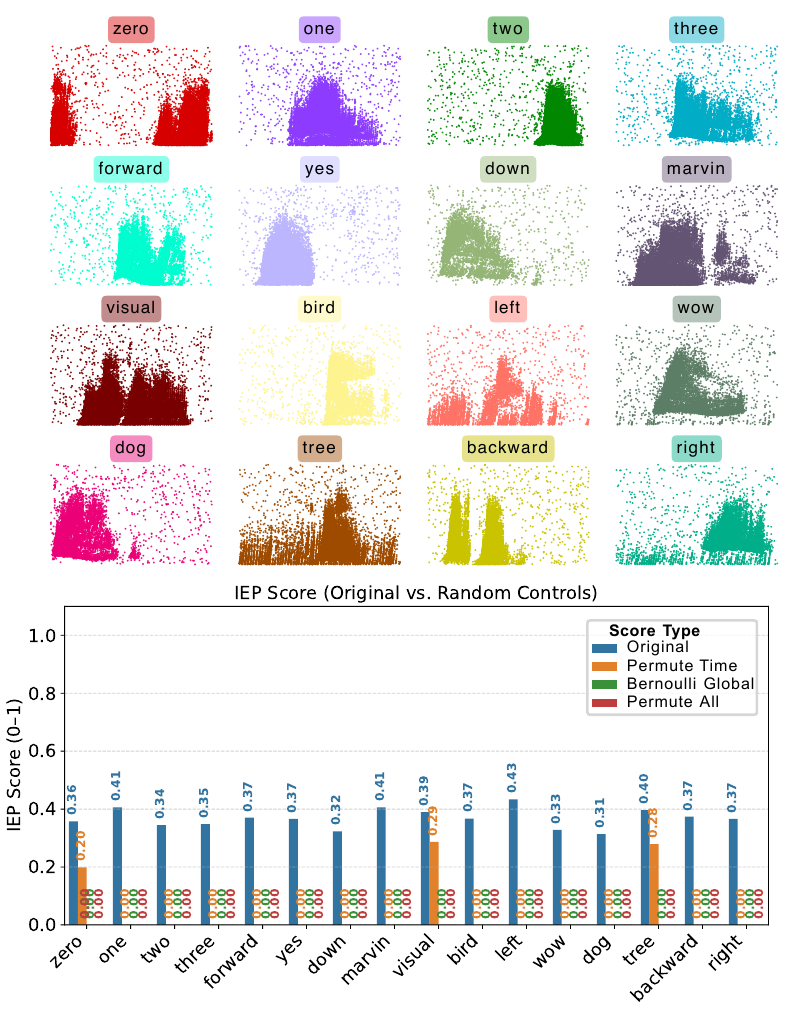}
  \caption{\textbf{$IEP$ on auditory-driven spike rasters from the Spiking Speech Commands (SSC) dataset.} 
\emph{Top}: representative 700-channel spike rasters for 16 spoken-word categories, generated by the 
\texttt{Lauscher} artificial cochlea model. 
\emph{Bottom}: $\IEP$ scores for each original raster (\textcolor{blue}{Original}) compared to the three surrogate controls introduced earlier: 
\textcolor{orange}{Permute Time}, 
\textcolor{green}{Bernoulli Global}, and 
\textcolor{red}{Permute All}.  
Original auditory rasters achieve higher $\IEP$ scores than their corresponding surrogates, reflecting the underlying spectrotemporal structure across words and the cochlea-generated spike trains.
See Table~\ref{tab:hyperparams_datasets} for the dataset-specific hyperparameters.
}
\label{fig:iep_audio}
\end{figure}

We next applied $IEP$ to artificial-cochlea-generated, high-dimensional spike trains from the Spiking Speech Commands (SSC) dataset \cite{cramer2020heidelberg}. SSC is derived from the Google Speech Commands dataset and converted into event-based representations using the Lauscher artificial cochlea model, yielding 700-channel spike rasters that reflect the spectrotemporal structure of each utterance. For consistency with our earlier benchmark, we selected one representative sample from the following 16-word categories (\textit{`zero', `one', `two', `three', `yes', `down', `marvin', `bird', `left', `dog', `tree', `backward', `right', `visual', `wow', `forward'}) and converted the spike times into binary channel-time matrices at the native 20~kHz resolution, which were subsequently downsampled to 2~kHz (0.5~ms bins) using a max-pooling operation over non-overlapping 10-sample windows. Example rasters are shown in the top panel of Fig.~\ref{fig:iep_audio}, illustrating characteristic speech-driven structure such as onset bursts, energy concentration in specific channels, and nonuniform temporal envelopes.

The bottom panel summarizes $IEP$ values for each original raster together with the same three surrogate controls introduced in the previous subsection. Across words, the original signals show consistently elevated $IEP$ relative to the surrogate groups, consistent with their pronounced spectrotemporal structure. 

Overall, these results show that the $IEP$ method can be applied beyond synthetic fractal benchmarks and remains sensitive to structure in realistic, stimulus-driven spiking data, even when that structure is subtler and embedded in high-dimensional auditory representations.

Finally, we note that among the controls, only \textit{Permute time} occasionally yields non-negligible $\IEP$ values, particularly in the Fractal-driven spike rasters. 
This occurs because this surrogate preserves the firing rate of each individual channel independently. 
In signals where the structure is dominated by spatial heterogeneity—such as specific channels being persistently active while others are silent (e.g., the horizontal banding in the flat/spatiotemporal sequences)—this preserved spatial variance can be sufficient to generate a positive score. 
In contrast, the \textit{Bernoulli global} and \textit{Permute all} controls destroy both spatial and temporal correlations, consistently reducing the signal to unstructured noise with effectively zero $\IEP$.

\section{How do different drivers of intelligence interact?}

Above, we have discussed the free-energy-minimization driver associated with FEP and provided details for a new driver for maximizing information enthalpy in a system. We also propose that these systems interact with a mechanism that attempts to keep neural systems engaged in information environments in a critical regime. Much like an n-body problem, we postulate that the state of a neural system is determined not by the sum of individual drivers, but by the complex, non-linear interplay of these influences within a high dimensional state-space. We distinguish our proposed informational n-body framework from established biophysical accounts of brain function, such as hierarchical multi-scale models~\cite{Robinson_2005_Multiscale,Robinson_2003_Neurophysical} or neural mass models. While existing models focus at detailing the physiological machinery, they do not intrinsically explain why or when a specific driver would influence neural-subsystems to adopt states that would prioritize one specific policy over another relative to the information environment that system is embodied within. Conversely, the framework presented below focuses on how the normative, information-theoretic imperatives presented above operate over this biological hardware. We view these perspectives as highly complementary: the biophysical bounds mapped by hierarchical and neural mass models effectively support the ``informational state-space'' through which our proposed informational drivers push a neural system, effectively navigating the information physics behind each specific subsystem. 

Future work will map these drivers into a more detailed biophysical model (detailed work which is beyond the scope of this paper). Instead, here we provide a summary of why this model is conceptually coherent and ideal to consider the different informational drivers already described in this paper which are proposed to shape biological neural intelligence. In doing so, we aim to capture two distinct aspects of neural systems that complicate straightforward hierarchical models of the brain. 1) That the brain consists of multiple interrelated interacting physical structures that engage in information processing in distinct but not entirely hierarchical ways. 2) That the brain operates at multiple timescales and these influence behavior. Given that network level activity alone is insufficient to fully explain behavior, we propose that the optimization for certain qualities of information structure better explains the dynamic system properties of brains that can range from oscillatory to chaotic. In particular we argue that external information with these specific properties can push systems into chaotic or stable regimes that can interact with a range of attractor dynamics known to persist in the brain.

\subsection{The brain consists of multiple interrelated interacting physical structures that operate across timescales.}

Neural systems possess multiple interrelated interacting and nested structures that are likely involved in intelligence, all connected with distinct geometry and topology. For the illustrative purposes here, we will consider dendrites, soma, and networks of cells with their own geometry and topology (although many other divisions and structures could also be proposed). When considered as separate systems, all operate simultaneously as systems of competing and cooperative dynamic functions in a loose hierarchy with some level of nesting. This can be understood by examining how each individual type of structure may process and transmit information both within and between each system. 

Dendrites extend from neurons in complex branching patterns and give rise to synapses typically on spines arising along the dendritic tree\cite{yuste_dendritic_2010}. It has long been established that dendrites will filter synaptic events as they propagate towards the soma\cite{magee_dendritic_2000, spruston_pyramidal_2008}. To achieve this, it has been proposed that dendrites can be modeled as multiple functional independent compartments which integrate different streams of information\cite{petreanu_subcellular_2009, spruston_pyramidal_2008, yuste_dendritic_2010}. That dendrites are capable of their own discrete electrophysiological activity in response to varied stimuli further supports a more complex role of these structures\cite{gidon_dendritic_2020}. Despite this complexity inherent in dendrites, these computations are then transmitted to the neural soma. 

At the somatic level, integration of synaptic inputs from dendrites can be modulated by numerous factors, including cell morphology, ion-channel expression and influences from supporting cells such as glia and astrocytes, leading to intricate regularization of neuronal dynamics that occur independently from dendritic filtering~\cite{magee_dendritic_2000}. In the cerebral cortex, at least seven excitatory and 16 inhibitory neurons have been identified using single cell RNAseq, along with many support glial cell types~\cite{johnson_cerebral_2017, zeisel_cell_2015}. Other methods have identified at least 49 transcriptomic types includes 23 types of inhibitory GABAergic, 19 excitatory glutamatergic and 7 non-neuronal cell types~\cite{tasic_adult_2016}. Regardless the exact number of cell types, it is clear that multiple exist and that different cell types will have both shared and distinct functionality that are linked to specific cellular morphology. However, it has been shown that morphology cannot completely account for regularizing dendritic input, indicating significant contribution by other cellular properties~\cite{magee_dendritic_2000}. In addition to changes in somatic morphology, the somatic expression of various ion channels can be dynamically varied.\cite{goaillard_ion_2021}. As such, the influence of neuronal somatic changes on the network cannot be well explained by dendritic influences alone justifying considering somatic properties as a separate (albeit related) system. For example, differences in somatic and dendritic inhibition have been identified to have diverse impacts on the activity and likely computation in the hippocampus\cite{miles_differences_1996, royer_control_2012}. 

While it may be tempting to consider networks as simply a sum of the lower-level parts described above, there is good reason to position networks of activity as a distinct operational system within the brain. The non-linear results of populations of interconnected neurons produces results not fully explainable by the sum of the parts and can display higher order dynamics\cite{ebitz_population_2021, saxena_towards_2019, schneidman_weak_2006}. Of particular note however, when considering networks, it is critical to consider both topology and geometry. Undoubtedly the topology of neural networks has a critical role in how information is processed and intelligence arises. For instance, computational models that enforce a comparable structure to that observed in cortical columns are more computationally efficient than the same sized network that is randomly organized \cite{haeusler_statistical_2007}. Canonical microcircuits have been proposed as a way to conceptualize the basic network elements that may underpin basic information processing that leads to intelligence\cite{bastos_canonical_2012}. It has been proposed that these canonical microcircuits function to facilitate predictive coding that can be detected through a network based on differing frequencies that arise through the non-linear processing of predictions and prediction errors\cite{bastos_canonical_2012}. Network learning rules have also been reflected in the relationship between different networked structures in a manner distinguishable from single neuron tuning functions\cite{ebitz_rules_2020}. Different types of processing in the network may operate under different rhythms, with some evidence that bottom-up signaling utilizes theta and gamma rhythms, while top-down signaling uses beta rhythms\cite{bastos_visual_2015}. This differential signaling constrains a functional hierarchy that may be changing dynamically over time in a context dependent manner to serve differential communication requirements through the networks. Yet while network activity is related to anatomical hierarchy, it is not necessarily constrained by it  \cite{markov_anatomy_2014}. Some analysis has suggested that geometry constraints may play a more fundamental role than the topological connectivity within the brain\cite{pang_geometric_2023}. However, it is plausible that the Exponential Distance Rule on eigenmodes (resonances within the system) inspired from Neural Mean Theory offers the core predictive utility in this approach as way of modeling the local influence of underlying dynamics from dendritic and somatic systems. However, despite these findings, network level activity remains insufficient to fully explain behavior. In some cases, the high redundancy of information in population wide activity means that single neuron activity can be a remarkable predictor of behavior as the signal component is conserved while the noise component can increase with additional cells\cite{pitkow_how_2015}. More stark examples can be seen in simpler organisms where the activation of specific single neurons can directly trigger complex behavior such as locomotion and escape behaviors \cite{carreira-rosario_mdn_2018, zhao_two-layer_2024, hecker_removing_2020}. As such, while there are multiple physical levels of systems that process information in the brain, a straight forward hierarchy does not best explain these dynamics.  

In addition to distinct physical systems, the brain also operates a different timescales which in turn modulate the overall system level impact these subsystems might produce given a series of inputs\cite{robinson_multiscale_2005,cavanagh_diversity_2020}. For example, in some cases data suggests that the functional architecture of the brain may be primed to operate on key time scales for activities such as attentional modulation \cite{helfrich_neural_2018}. More broadly, it has been proposed that the temporal flexibility of information flows in the brain facilitates adaptive behavior while responding to dynamic environments - an evolutionary requirement for any animal\cite{cavanagh_diversity_2020}. At the network level in the cortex, a hierarchy of intrinsic timescales has been identified across the cortex that may link neurophysiological properties to functional specialization \cite{murray_hierarchy_2014}. Moreover, different network structures can show multiple timescales of information processing, such as in the hippocampus and cortex, that nonetheless permits adaptive cooperative interactions \cite{tang_multiple_2021}.  Yet the importance of timescales persists across different levels of a neural system described above, with individual cells and even dendritic networks encoding information with time based constants \cite{cavanagh_diversity_2020}. For instance, in studies of working memory it has been shown that cells with short timescales appear to carry memory information relatively early during encoding, while longer timescale activity in cells predicted during information processing that occurred later \cite{wasmuht_intrinsic_2018}. Overall, the different intrinsic neural timescales of different systems has been well established in processing information received by the system, both in terms of integrating and segregating signals \cite{golesorkhi_brain_2021}.   Given the importance of timescales and different rates of information processing at different sub-systems of the brain, it can be argued that it is insufficient to only model neural systems in a hierarchical fashion that accounts for the anatomical organization, but not the physiologically relevant temporal encodings. Any model of how neural system function leads to intelligence must take into account the time based changes which occurs both within and between the different subsystems. 

\subsection{The interaction of systems across timescales supports adopting n-body-inspired system modeling}

As established above, neural systems have multiple sub-systems each with their own time-based oscillatory dynamics which modulate the interaction between and within subsystems. While a hierarchy undoubtedly exists in how these subsystems may influence each other and the overall system behavior, the hierarchy is not strict as it is established that in specific cases even single neuron activity can strongly predict an organism's behavior. When considering mathematical models for systems with sub-components, each of which have their own dynamics while affecting other sub-components across time, the justification for proposing that the class of equations for investigating n-body problems is the ideal way to model a system like the brain. n-body problems are infamously hard to solve as a general solution is computationally intractable. These equations can exhibit rich dynamics that range for stable oscillatory patterns, to those that are considered chaotic. This breadth of behavior is consistent with the high dimensional states that biological neural systems can transverse. Yet unlike celestial objects which do not show any observable evolutionary drivers to actively maintain their system, biological neural systems have evolved to maintain certain imperative drivers. Here we argue that an interaction between the internally processes inherent across spatial and temporal levels in the brain, coupled with different information drivers we described above, is best modeled as an n-body formulation. Under this framework, the ``motion'' of different neural systems progressing through different states have dynamic interactions with the other neural systems. It is only the inherent biological predisposition to the information based drivers we described above that constrain the system in such a way as to allow the range of intelligent behaviors arise out of what would otherwise be a system that would progress to either fully chaotic regimes or stagnant ones. 

While an unknown (n) number of drivers could play a role here and across multiple biophysical systems, below we illustrate the concept by discussing the 3 concepts emphasized in this paper so far: effects across information enthalpy maximization, free-energy minimization and a drive to remain in a critical regime while processing information.

\subsection{Relationship between Information Entropy, Information Enthalpy and Neural Criticality}

\begin{figure}[t]
    \centering
    \renewcommand{\thesubfigure}{\Alph{subfigure}} 
    \captionsetup[subfigure]{
        position        = top,
        labelsep        = none,
        justification   = raggedright,
        singlelinecheck = false,
        skip            = 0pt,
        labelfont       = bf,
    }
    
    \begin{subfigure}[b]{\textwidth}
        \caption{}
        \label{fig:assisted_dynamics}
        \includegraphics[width=\linewidth]{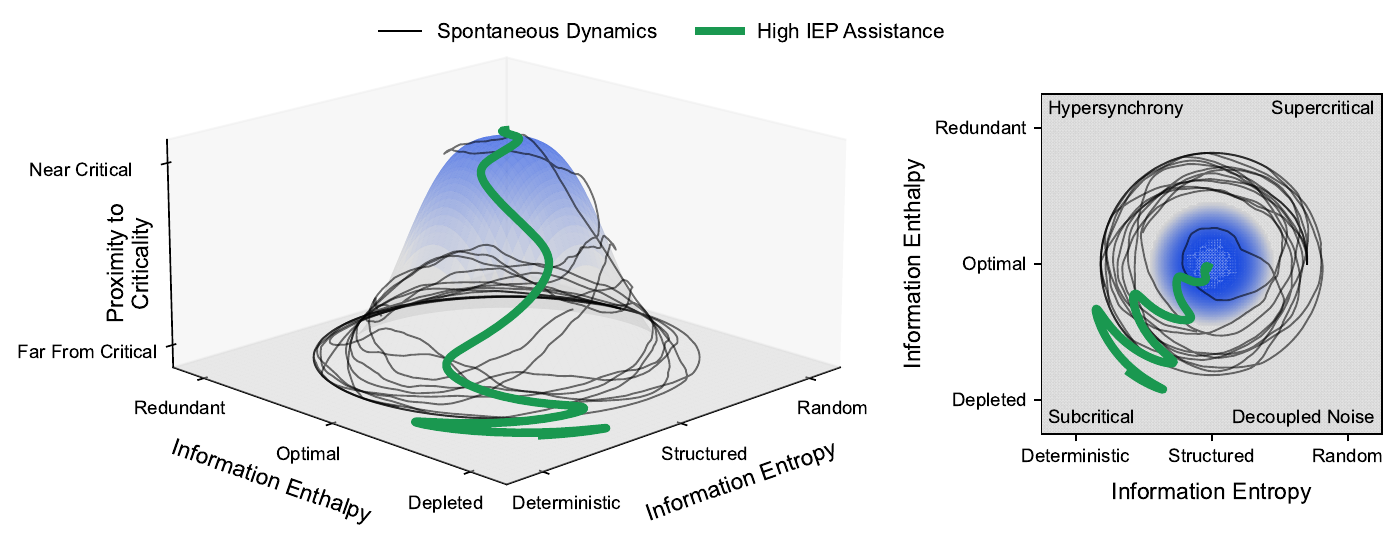}
    \end{subfigure}
    \vspace{0.5em}
    \begin{subfigure}[b]{0.49\textwidth}
        \caption{}
        \label{fig:info_entropy_levels}
        \includegraphics[width=\linewidth]{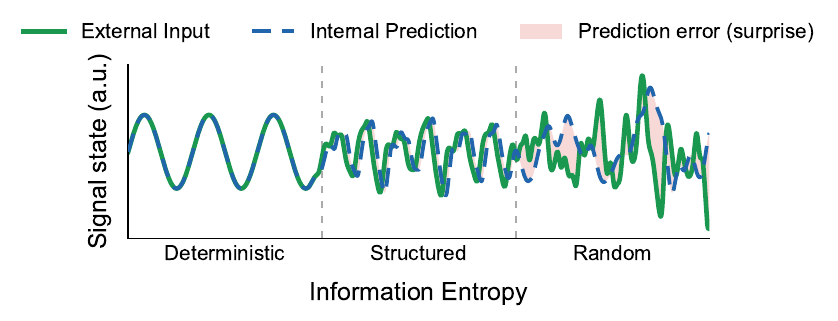}
    \end{subfigure}
    \begin{subfigure}[b]{0.49\textwidth}
        \caption{}
        \label{fig:info_enthalpy_levels}
        \includegraphics[width=\linewidth]{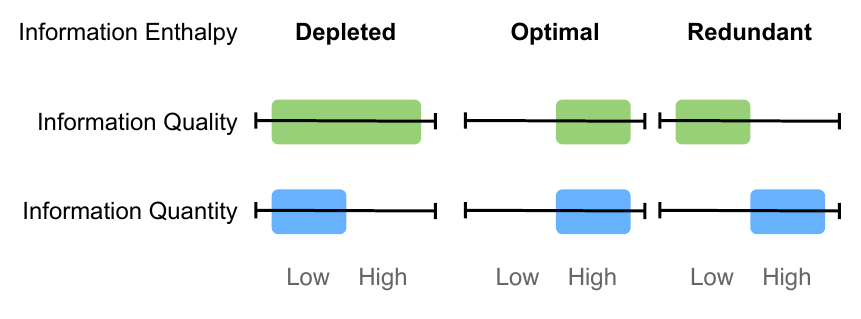}
    \end{subfigure}
    \caption{
    Schematics illustrating \textbf{(A):} dynamical interaction between Information Enthalpy, Information Entropy and Criticality Potential of a neural information processing system. The black trajectory represents Spontaneous Dynamics that explore the landscape without external guidance. In contrast, the green trajectory shows how structured inputs (i.e. with high IEP) can drive the system closer towards criticality, despite the inherent instability of critical states; 
    \textbf{(B):} different levels of information entropy and the relationship between the states of external input and internal predictions; and 
    \textbf{(C):} different levels of information enthalpy and their relative composition in terms of information quality and quantity.
    }
\end{figure}

We posit that the state of a neural system can be mapped through the continuous interplay between information entropy and information enthalpy. To understand this dynamic landscape, we first consider information entropy through two complementary lenses (see~\Cref{fig:info_entropy_levels}). From the perspective of the external signal, information entropy quantifies the uncertainty associated with its possible states. From the neural system's perspective, entropy reflects the degree of expected surprise associated with the input signal and, consequently, the prediction error that can arise when the input deviates from the system's internal predictions. We associate low entropy with input signals with minimal complexity, leading neural systems to develop \textit{deterministic} internal representations with minimal prediction error. Conversely, high entropy signals, such as \textit{random} signals, lack predictable structure, preventing neural systems from building reliable internal models. We hypothesize that some degree of prediction error, such as with complex but \textit{structured} signals, is necessary to drive neural adaptation, while remaining within the system's available adaptive capacity.

We also characterize information enthalpy, defined as the predictive-information resource capacity of the system, by the balance between information quality and information quantity (see~\Cref{fig:info_enthalpy_levels}). We associate the quality of information with its \textit{usefulness}, being the degree to which information contained in, for instance, the state $\Mn{t}$ is predictive of future states $F$.
A neural system could be referred to as having \textit{depleted} predictive-information resources when its predictive-information capacity (information enthalpy) is insufficient to enable the system to acquire sufficient predictive information to develop robust internal predictive models. In contrast, an abundance of predictive-information capacity could lead neural systems to make \textit{redundant} resource commitments, for instance by using energetic or storage resources to maintain predictive information that is already represented, and therefore provides little additional predictive utility. Under this model, a regime where information enthalpy is at capacity and additional resource is redundant, the underlying driver to maximize information enthalpy is minimized and other drivers are proportionally more relevant. Therefore, we hypothesize that a neural system is in an \textit{optimal} state of information enthalpy when its predictive-information capacity is sufficiently large and the information supporting this capacity is predominantly high-quality and non-redundant.

Beyond this relationship, there also exists a coherent association in which critical dynamics can be viewed as a responsive and resilient operating regime for neural systems. Critical dynamics are typically associated with long-range spatiotemporal dynamics~\cite{klaus_statistical_2011}, a broad dynamic range that enables responsiveness to changes in input~\cite{shew_neuronal_2009, kinouchi_optimal_2006}, and heightened susceptibility to structured input from external environments~\cite{habibollahi_critical_2023}. Rather than corresponding to minimal entropy states, critical regimes combine variability with long-range organization, placing them in an intermediate regime between highly ordered and highly disordered dynamics~\cite{hesse_selforganized_2014}.

 The information enthalpy conceptualization accords naturally with this view. Information enthalpy, $\mathcal H_I$, represents the predictive-information resource capacity available to the system under its information-processing, signal-dependent, and energetic constraints. We hypothesize that maintaining a sufficiently high information enthalpy provides the system with the capacity to acquire and retain useful predictive information about its environment, thereby supporting the complex internal dynamics required to remain near criticality. Importantly, this requires not simply an abundance of information, but information of sufficiently high quality to support prediction without being predominantly redundant or uninformative. A depleted information enthalpy regime may therefore provide insufficient predictive-information capacity to sustain critical dynamics, whereas an abundance of low-quality information may commit finite information-processing and energetic resources without providing a corresponding increase in predictive-information capacity. We hypothesize that an optimal information enthalpy regime, characterized by a sufficiently high predictive-information capacity supported by abundant, high-quality and non-redundant information, provides favorable conditions for maintaining critical-like dynamics.

In Figure~\ref{fig:assisted_dynamics}, we illustrate the dynamical interaction between information entropy, information enthalpy, and proximity to criticality. This landscape defines several discrete states: (1) Subcriticality, which represents a state of low dynamical richness akin to a quiescent or rigid network where deterministic entropy and depleted enthalpy prevent complex information processing, (2) Supercriticality, which resembles runaway excitation or chaotic activity where random entropy and redundant enthalpy overwhelm the system's capacity for stable organization, (3) Hypersynchrony, characterized by highly repetitive and synchronized firing patterns that represent deterministic entropy and redundant enthalpy, (4) Decoupled Noise, which reflects a state of stochastic and uninformative activity characterized by random entropy and depleted enthalpy, and (5) a central ideal near-critical regime that is best suited for neural information processing but also remains unstable. Spontaneous neural dynamics, represented by the black trajectory, continuously explore this landscape, balancing homeostatic biological mechanisms (providing upward pressure) against constant perturbations of sensory input and metabolic fatigue (downward pressure). 

We therefore hypothesize that external input with high Information Enthalpy Potential (IEP) can act as a driver that helps stabilize neural dynamics near the critical regime. As illustrated by the green trajectory in~\Cref{fig:assisted_dynamics}, a structured input (high IEP) drives the system toward a trajectory close to optimal information enthalpy and structured entropy, guiding the system from a state far from criticality toward the critical peak. We propose that such informational assistance may also influence functional connectivity and facilitate the emergence of complex adaptive behaviors. In future work, we aim to empirically investigate the impact of high-IEP input on neural dynamics to empirically evaluate these theoretical predictions.


\section{Discussion}

This paper attempts to tackle one of the most ambitious scientific questions still facing humankind: what leads a collection of cells to be able to self-organize, adopt states, and persist across decades in such a way as to display general intelligence potent enough to reshape the environment that gave rise to it? Or more simply: how does intelligence arise from a brain? In asking this question we face the paradox proposed nearly 90 years ago by Emerson M. Pugh: \textit{``If the human brain were so simple that we could understand it, we would be so simple that we couldn't''}. Naturally this paper does not answer these questions fully. However, we believe it establishes a conceptual stepping stone. The ideas here can be tested by combining advancements across mathematics, computing, biology and bioengineering, electrical engineering, and emerging fields like Synthetic Biological Intelligence (SBI) that seek to leverage simplified neural systems for information processing and intelligence. If the theory presented here has merit, key falsifiable implications have already arisen. For the sake of brevity, we make explicit four key examples here:  1) Increasing the $\IEP$ of sensory input, while holding firing-rate statistics approximately fixed, should move neural dynamics toward canonical signatures of criticality, up to saturation imposed by capacity and biological costs; 2) The sensitivity of criticality metrics to changes in input structure should peak near the boundary between ordered and disordered regimes; 3) Reductions in effective throughput capacity $\Cr{t}$ should weaken both information enthalpy accumulation and the stability of critical-like dynamics, even under highly structured input; and 4) The n-body formulation should provide a more effective physics-based constraint to shape the neurocomputational dynamics of biological neural networks when the information entropy and information enthalpy of input signals is constrained in particular ways. Some of these implications are relatively simple to test and others require additional technological advancement to confidently explore. 

An implicit aspect of the information enthalpy formulation presented here was that this is proposed as a complementary and interacting normative principle for active inference with that of the FEP. Many of the same properties that make active inference compelling as a theoretical basis become stronger by the addition of a second normative principle that potentially operates via different neural mechanisms. The broad processes of actively sampling and predicting the environment and associated outcomes of actions remain; however, based on the requirements of the neural system carrying out these processes, what the actions or predictions will aim to optimize will vary. This simplifies accounts in FEP literature that currently require complex generative models to explain behavior. Arguably however, this model would better accord with earlier conceptions that suggested aversions to ``dark rooms'', self-induced instabilities (sometimes called ``autovitiation''), and similar features, were essentially due to predetermined genetic priors. Here, these genetic priors are made explicit, but under a separate and competing driver from that described in the FEP. In this case, the priors are established by the driver to maximize the amount of usable information that a system collects. Future work will have to further explore this relationship, especially through well controlled experimentation.  

Finally, while this paper makes great efforts to introduce in detail the information enthalpy driver and to contrast this new theoretical concept against the established FEP-based free-energy minimization concept, we do not propose that other drivers do not exist. This is why we introduce the notion that there is an inherent driver for criticality in neural systems, as previous work suggests this is a self-organizing feature of dynamic systems embodied within a structured information landscape where information storage and transfer is a requirement \cite{habibollahi_critical_2022}. What is likely however, is that there are primary and secondary drivers to neural systems. In this case, we would propose information enthalpy maximization and free-energy minimization are primary drivers, while neural criticality maintenance are secondary drivers. Functionally, this would work similarly to an astrophysics n-body formulation, where the impact of individual bodies on the whole system will differ based on the mass of each system, due to the different gravitational forces.  The n-body-inspired framework presented here aims to provide a conceptualization for systems with distinct drivers may interact to produce changes in neural state spaces as a result of conflicting forces, even if detailing exact biophysical system level interactions that might drive the system as a whole is beyond the scope of this (already extensive) manuscript. Fortunately however, the benefit in the approach here is that, under controlled conditions, the dynamics of these different drivers as they may apply to different biophysical system can be measured, manipulated, and the theoretical perspective then updated. By exploring across scales and complexity, this iterative approach offers the best chance to understand how the brain works to give rise to intelligence, or at the very least, that of a simple brain.

\subsection*{\textbf{Acknowledgments}}
This work is supported by the U.S. Defense Advanced Research Projects Agency (DARPA). The views, opinions, and/or findings expressed are those of the author(s) and should not be interpreted as representing the official views or policies of the Department of Defense or the U.S. Government. The authors thank Professor Zdenka Kuncic, A/Professor Hamish Meffin, Dr Michael Barros, and Andy C. Kitchen for insightful discussion, and Dr Nicole Kerlero de Rosbo for proofreading.

\subsection*{\textbf{Declarations}}
The authors declared the following potential conflicts of interest with respect to the research, authorship, and/or publication of this article: B.J.K., J.Z., F.H. and V.B. are employees of and may hold shares or another interest in Cortical Labs, a research-focused start-up working in a space and holding patents related to this article. No specific incentive was provided to any author for contribution to this article.  

\subsection*{\textbf{Data and Code Availability}}
Data and code will be made available at a publicly accessible repository for further investigation and reproducibility after the manuscript finishes peer-review.


\bibliography{references,references-2}  


\appendix


\section{Appendix }
\subsection{Information-storage dynamics and historical capacity depth }
\label{appsec:T_cap}

The system's information storage, or cache, can be modeled as a fading memory, similar to a simple leaky reservoir or varying capacity, or a resistance-capacitance (RC) circuit with varying resistance. This can be modeled using a linear time-varying difference equation:
\begin{align}
    \Cst{t}
=
    \big(
        1
    -
        \alpha_t
        \Delta t 
    \big)
    \Cst{t-1}
+
    \Delta t 
    \,
    \Cr{t-1}  
\label{eq:CstoreExtended}
\end{align}
where \(\alpha_t = \Tcap{t}^{-1}\) is a time-varying ``leakage'' factor. $\Tcap{t}$ can therefore be interpreted as \textit{historical capacity depth} of the information on sensory input $O$ and internal states $X$ that the system can actively maintain, that is, a changing parameter that controls the non-Markovian history depth). This equation is coupled with the difference equation for $\Cr{t}$
\begin{align}
    C_{\mathrm{r},t}
=
    \left(
        u_t - v_t
    \right)
    \Delta t
+
    \big(
        1
    -
        \Delta t\delta_{\mathrm{decay}}
    \big)
    \,\Cr{t-1} 
\end{align}
where $u_t$, $v_t$, and $\delta_{\mathrm{decay}}$ are, respectively, the capacity-build rate, capacity-pruning rate, and passive capacity-decay rate if not maintained. To avoid resource depletion, we also introduce a minimum baseline retrospective window $t_\mathrm{cap}$, 
\begin{equation}
    \Tcap{t} 
=
    \max
        \left[
            t_{\mathrm{cap}}, 
            \frac{\Cst{t}}{
                \Cr{t}
            - 
                \Delta \Cst{t}/\Delta t}
        \right]
\label{eq:TcapComplete}
\end{equation}
Here, $t_\mathrm{cap}$ prevents a zero-horizon state for the case $\Cst{t} \rightarrow 0$, allowing the system to continuously bootstrap its predictive generative models. This is particularly relevant in the context of active inference, where this retrospective window serves as the mandatory informational baseline for prospective action selection. Because future policy optimization $\pi$ and expected information enthalpy maximization require conditioning on historical trajectories to preserve predictive consistency, the system’s prospective predictive horizon is strictly bounded by its retrospective retention capability. 

In our implementation, we establish a strict structural constraint: the neuronal system cannot project its future policies deeper than the historical context it can physically retain. Mathematically, this maps to a path-dependent conditional optimization where the prospective horizon length $H$ is fundamentally bounded or scaled by the retrospective capacity: $H \le T_{\mathrm{cap},t}$

If we introduce a fixed structural ceiling of the storage substrate $C^\mathrm{max}_{\mathrm{st}}$, that caps the maximum value of $\Cst{t}$, the dynamic equation~\eqref{eq:CstoreExtended} can be modified to include this term as
\begin{align}
    \Cst{t}
=
    \left(
        1
    -
        \alpha_{t-1}
        \,
        \Delta t
    \right)
    \Cst{t-1}
+
    \Delta t 
    \,
    \Cr{t-1}
    \left(
        1
    -
        \frac{\Cst{t-1}}{C^\mathrm{max}_\mathrm{st}}
    \right).
\end{align}
This allows us to introduce an effective throughput capacity 
\begin{align}
    C^\mathrm{eff}_{\mathrm{r},t}    
=
    \Cr{t}
        \left(
        1
    -
        \frac{\Cst{t}}{C^\mathrm{max}_\mathrm{st}}
    \right),
\end{align}
that, for $\Cst{t} \rightarrow C^\mathrm{max}_\mathrm{store}$, tends to zero, preventing overflowing of the storage capacity. The retention capacity depth $\Tcap{t}$ still follows equation~\eqref{eq:TcapComplete} where $\Cr{t}$ is now replaced with $C^\mathrm{eff}_{\mathrm{r},t}$.

\subsection{Information rate}
\label{appsec:info-rate}

The long-term average rate of the directed information~\cite{tsur_directed_2026} can be defined as:
\begin{equation}
    \overline{\mathcal{R}}\big(P\!\to\! \cMn \big)
\coloneqq 
    \lim_{T\to \infty}
    \frac{1}{T}
    I\big(P^T\!\to\! M^{(n),T}\big),
\label{eq:long-term-DI-rate}
\end{equation}
whenever this limit exists. 

While the ITC is defined in the limit $T\rightarrow \infty$, in practice, the system is subject to the finite temporal bound $T_\mathrm{cap}(t)$. The long-term average rate at time $t$ can then be evaluated for $T \rightarrow T_{\mathrm{cap},t}$, 
\begin{equation}
    \overline{\mathcal{R}}\big(P\!\to\! \cMn \big)
\coloneqq 
    \frac{I\big(P^{T_{\mathrm{cap}}}\!\to\!M^{(n),T_{\mathrm{cap}}}\big)}{T_{\mathrm{cap}}}
\label{eq:DI-rate-finite-der}
\end{equation}
%

\subsection{Homeostatic penalty}
\label{appsec:homeostaticpenalty}

An example of homeostatic penalty function is:
\begin{equation}
    \Psi(h_t)
=
    \gamma\big[\sigma\log\!\big(1+e^{(h_t-h_{\mathrm{tol}})/\sigma}\big)\big]^p
\end{equation}
Here, $h_t$ is the homeostatic deviation, $h_{\mathrm{tol}}$ is the tolerance radius below which the penalty remains negligible, $p$ is the growth exponent determining how steeply the penalty grows once deviations pass the tolerance threshold $h_t>h_{\mathrm{tol}}$, and $\gamma$ is the penalty gain. The core formulation utilizes a softplus function $\log\!\big(1+e^{x/\sigma}\big)$ to provide a smooth yet increasing penalty as homeostatic deviations occur, which can be softened by increasing $\sigma$ as a smoothness parameter.

\subsection{Setup and symbolization}
\label{appsec:sequence_setup}

While theoretically $\mathrm{IEP}$ score may be computed on information of any dimensionality, to provide a workable example we consider a 2D binary raster (channels x time), denoted by $X\in\{0,1\}^{C\times T}$, where $\mathbb{T}=\{0,1,\dots,T-1\}$. As will be seen in the examples in section~\ref{sec:IEP-dem}, this is convenient as a way to deliver a complex signal through patterned set of stimulation such as may be delivered via a microelectrode array to neural cells. 

\begin{figure}[ht]
  \centering
  \includegraphics[width=0.6\textwidth]{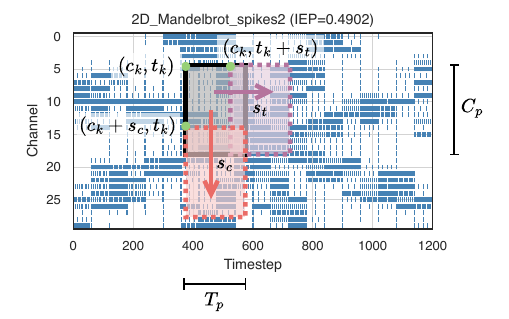}
  \caption{\textbf{Schematic description of preparation of patches.} Patches of size $(C_p, T_p)$ are created from coordinate sets $\mathcal{I}_c$ and $\mathcal{I}_p$ by moving with strides $s_c$ and $s_t$ respectively, with the $k$-th patch (shown in black) initiating from coordinate $(c_k, p_k)$.}
  \label{fig:symbolizationAppendix}
\end{figure}

To evaluate the $\IEP$ at specific scales, the raster $X$ is partitioned into \textit{patches} of size $C_p \times  T_p\in\bbN^2$, that can be thought of as characterizing the scale of observation. We also define the \textit{strides} $s_c,s_t\in\bbN$, to iterate over the patches. Specifically, each patch $W_k$ is constructed as follows:
\begin{enumerate}
    \item \textbf{Coordinate Sets:} We define the set of starting indices $\mathcal{I}_c$ and $\mathcal{I}_t$ in terms of the chosen strides $s_c$ and $s_t$:
    \begin{equation}
    \mathcal{I}_c:=\{0,s_c,2s_c,\dots,C-C_p\},\qquad 
    \mathcal{I}_t:=\{0,s_t,2s_t,\dots,T-T_p\}.
    \end{equation}
    This provides a collection of start-index pairs $(c_k,t_k) \in \mathcal{I}_c \times \mathcal{I}_t$, where $1\leq k \leq N $ is the patch index, ($N$ total number of patches). 
    \item \textbf{Patch extraction:} Each patch $W_k\in \{0,1\}^{C_p \times T_p}$ is extracted from $X$ using the rows $c_k:(c_k+C_p-1)$ and columns $t_k:(t_k+T_p-1)$. 
\end{enumerate}
Notably, one can set $C_p \geq s_c$ and $T_p \geq s_t$ to ensure patch overlap, thus preserving spatial and temporal correlations across patches (see Fig.~\ref{fig:symbolizationAppendix}), or to avoid this if desirable. This flexibility is necessary given the variety of scales at which information structure may exist. 

From these patches, we generate a \textit{block-occupancy map}, with block size $(p_c,p_t)$ and threshold $\theta\in[0,1]$. The parameters $p_c$ and $p_t$ are selected to appropriately aggregate a specific number of channels and time steps, respectively. The block occupancy pooling operation is then defined as:
\begin{equation}
    \mathsf{Pool}_{\theta,p_c,p_t}:\{0,1\}^{C_p\times T_p}\to\{0,1\}^{\lfloor C_p/p_c\rfloor\times\lfloor T_p/p_t\rfloor}. 
\end{equation}
A pooled bit is assigned a value of $1$ only if its block-occupancy fraction is greater than or equal to  $\theta$. 

Each pooled patch is transformed into a binary \textit{token} $s_k$ according to
\begin{align}
    s_k=\varpi \big(\mathsf{Pool}_{\theta,p_c,p_t}(W_k)\big).
\end{align}
where the \textit{packing map} $\varpi : \{0,1\}^{m\times n}\to \{0,1\}^L$ flattens the pooled grid into a binary sequence $s_k$. By aggregating these tokens from the full collection of patches $\{W_k\}_{k=1}^N$, we construct the sequence $S = (s_1, \dots, s_N)$. The alphabet $\mathcal{A}$ is then defined as the set of all unique tokens $\sigma$ present within this sequence:
\begin{equation}
    S=(s_1,\dots,s_N), 
\qquad 
    \mathcal{A} = \{\sigma \mid \sigma \in S \},
\end{equation}
The alphabet cardinality $M = |\mathcal{A}|$ is not fixed in advance, but it is a data-dependent parameter that represents the variety of unique patterns present in the signal at the chosen scale.
This makes the pooling and tokenization steps important for assessing how structured, non-degenerate, and persistent across scales the resulting token sequence is at a given level. This approach provides convenient hyperparameters that can be modified and explored to ensure appropriate granularity of analysis.

\renewcommand{\thesection}{S\arabic{section}}
\renewcommand{\thesubsection}{S\arabic{section}.\arabic{subsection}}
\renewcommand{\thetable}{S\arabic{table}}
\renewcommand{\thefigure}{S\arabic{figure}}

\section{Supplementary Methods and Implementation Details}
\label{sec:supp_methods}

\subsection*{IEP demonstration - Hyperparameter settings by dataset}
\label{sec:supp_hparams}
The hyperparameters used for the fractal and auditory demonstrations are summarized in Table~\ref{tab:hyperparams_datasets}. Dataset-specific settings were chosen to accommodate the markedly different spatial and temporal organization of the two signal classes. The synthetic fractal rasters contain geometrically distributed structure spanning multiple spatial and temporal scales, whereas the SSC rasters exhibit more localized spectrotemporal organization, with temporally extended regions of dense activity concentrated within particular channel bands and comparatively sparse activity elsewhere. Accordingly, the SSC analysis used a longer temporal patch ($T_p=256$ versus $64$) to capture these extended speech-driven activity envelopes, while using smaller spatial patches ($C_p=4$ versus $8$). Setting the spatial stride equal to the patch size ($s_c=C_p=4$) provides non-overlapping coverage of the full channel axis, allowing localized channel-wise structure to be retained without redundant spatial overlap. The lower occupancy threshold ($\theta=0.4$ versus $0.5$) accommodates the greater sparsity surrounding the concentrated auditory activity.

The remaining parameters were adjusted consistently with this difference in signal organization. The lower entropy exponent for SSC ($\alpha=0.3$ versus $0.5$) reduces the down-weighting of lower-entropy tokenizations that can arise from localized, concentrated activity, while the lower mixing weight ($\beta=0.6$ versus $0.7$) places relatively greater emphasis on local entropy-weighted redundancy. The SSC analysis also uses the more compact block-scale set, emphasizing persistence across shorter and intermediate token scales, whereas the fractal analysis extends to substantially larger block sizes to capture its broader multiscale organization. PSC weighting was disabled for the SSC analysis because its pronounced, locally concentrated spectrotemporal structure was already directly captured by the tokenization, avoiding additional replication-based weighting of already prominent activity epochs; in contrast, PSC weighting was retained for the fractal rasters to emphasize structured regions within their more spatially distributed and multiscale organization.

\begin{table}[ht]
\centering
\small
\setlength{\tabcolsep}{8pt}
\begin{tabular}{lcc}
\toprule
\textbf{Hyperparameter (paper notation)} &
\textbf{Fractal-driven spike rasters} &
\textbf{Auditory-driven (SSC)} \\
\midrule
$C_p$ (spatial patch size)        & $8$   & $4$ \\
$T_p$ (temporal patch size)       & $64$  & $256$ \\
$s_c$ (spatial stride)            & $8$   & $4$ \\
$s_t$ (temporal stride)           & $1$   & $1$ \\
$p_c$ (spatial pool size)         & $2$   & $4$ \\
$p_t$ (temporal pool size)        & $2$   & $2$ \\
$\theta$ (occupancy threshold)    & $0.5$ & $0.4$ \\
\midrule
$\alpha$ (entropy exponent)       & $0.5$ & $0.3$ \\
$\beta$ (mixing weight)           & $0.7$ & $0.6$ \\
$\mathcal{B}=\{B_i\}$ (block scales) &
$\{5,10,20,40,80,160\}$ &
$\{1,2,4,8,16,32\}$ \\
$M$ (surrogate count)             & $60$  & $50$ \\
\midrule
$D$ (PSC embedding dimension)     & $5$   & - \\
$\tau$ (PSC ordinal delay)        & $3$   & - \\
$w$ (PSC window length)           & $60$  & - \\
$\lambda$ (PSC replication strength) & $1$ & - \\
$\nu$ (PSC normalization choice)  & \texttt{mean} & - \\
\bottomrule
\end{tabular}
\caption{Hyperparameters used for tokenization, entropy-weighted redundancy, multi-scale complexity, and PSC-based time weighting across the two datasets.}
\label{tab:hyperparams_datasets}
\end{table}

\end{document}